\documentclass{aa}
\usepackage{graphicx, epsfig, epsf, txfonts, natbib,epstopdf}
\usepackage{subfigure, tabularx, appendix, dcolumn,appendix}
\usepackage{color}
\usepackage[colorlinks=true,allcolors=blue]{hyperref}

\newcommand{\EQ}{\begin{equation}}
\newcommand{\EN}{\end{equation}}
\newcommand{\EQA}{\begin{eqnarray}}
\newcommand{\ENA}{\end{eqnarray}}

\def\kms{\hbox{km$\;$s$^{-1}$}} 
\def\arcsec{\hbox{$^{\prime\prime}$}}

\begin{document}

\title{Eruptions from coronal hole bright points: Observations and non-potential modeling}

\author{Maria~S. Madjarska\inst{1, 2} \and Klaus Galsgaard\inst{3} \and Duncan H. Mackay\inst{3} \and Kostadinka Koleva\inst{4} \and Momchil Dechev\inst{5}}

\offprints{madjarska@mps.mpg.de}
\institute{
Max Planck Institute for Solar System Research, Justus-von-Liebig-Weg 3, 37077, G\"ottingen, Germany
\and
Astronomy Program, Department of Physics and Astronomy, Seoul National University, Seoul 08826, Republic of Korea
\and
School of Mathematics and Statistics, University of St Andrews, North Haugh, St Andrews, KY16 9SS, Scotland, UK
\and 
Space Research and Technology Institute, Bulgarian Academy of Sciences, Acad. Georgy Bonchev Str., Bl. 1, BG-1113 Sofia, Bulgaria
\and
Institute of Astronomy and National Astronomical Observatory, Bulgarian Academy of Sciences, 72 Tsarigradsko Shose blvd., 1784 Sofia, Bulgaria}

\date{Received date, accepted date}

\abstract{We report on the third part of a series of studies on eruptions associated with small-scale loop complexes named coronal bright points (CBPs).}
{A single case study of a CBP in an equatorial coronal hole with an exceptionally large size is investigated to  extend our understanding of the formation of mini-filaments, their destabilisation and the origin of the eruption triggering the formation of jet-like features recorded in the extreme-ultraviolet (EUV) and X-ray emission. We aim to explore the nature of the so-called micro-flares in CBPs associated with jets in coronal holes and mini coronal mass ejections in the quiet Sun.}
{Co-observations from the Atmospheric Imaging Assembly (AIA) and Helioseismic Magnetic Imager (HMI) on board the Solar Dynamics Observatory, and GONG H$\alpha$ images are used together with a Non-Linear Force Free Field (NLFFF) relaxation approach, where the latter is based on a time series of HMI line-of-sight magnetograms.}
{A mini-filament (MF) that formed beneath the CBP arcade around 3--4~h before the eruption  is seen in the H$\alpha$ and EUV AIA images to lift up and erupt triggering the formation of an X-ray jet. No significant photospheric magnetic flux concentration displacement (convergence) is observed 
and neither is magnetic flux cancellation between the two main magnetic polarities forming the CBP in the time period leading to the MF liftoff. The CBP micro-flare is associated with three flare kernels that formed shortly after the MF liftoff. No observational signature is found for reconnection beneath the erupting  MF.  The applied NLFFF modelling successfully reproduces both the CBP loop complex as well as the magnetic flux rope that hosts the MF  during the build-up to the eruption.}
{}
\authorrunning{Madjarska et al.}

\maketitle

\section{Introduction}
\label{intro}

Coronal bright points (CBPs) have been intensively studied for almost five decades. They represent a set of small-scale coronal loops that connect magnetic flux concentrations of opposite polarity. As the plasma confined in these loops is heated to over a million degrees, they are seen with enhanced emission in EUV and X-ray. CBPs are found to be uniformly distributed in the solar corona of the quiet Sun, coronal holes and in the vicinity of active regions. This paper is the third of a series of studies that  investigate the eruptive behaviour of CBPs. \citet[][hereafter Paper~I]{2018A&A...619A..55M} explored the morphological and dynamical evolution of eruptions associated with CBPs in the context of their full lifetime evolution. The follow-up study by \citet[hereafter Paper~II]{2019A&A...623A..78G} employed data-driven modelling based on a Non-linear Force-Free Field (NLFFF) relaxation code to reproduce the time evolution of the magnetic field of these eruptions, and provided insight into the possible causes for destabilisation and eruption. An overview of the observational findings and modelling of CBPs and related phenomena are given in Papers~I and II. \citet{2019LRSP...16....2M} provides a detailed review on CBPs.

Here we briefly summarize the main findings on the eruptions from quiet Sun CBPs from Papers~I and II. Paper~I reports that 76\%\ of the studied CBPs (31 out of 42) hosted at least one eruption during their lifetime. The study then 
explored the observational properties of 21 eruptions associated with 11 quiet Sun CBPs. The eruptions occurred on average $\sim$17~h after the CBP formation, where the typical lifetime of CBPs in images taken with the Atmospheric Imaging Assembly (AIA) on board the Solar Dynamics Observatory (SDO)
in the Fe~{\sc xii}~193~\AA\ channel (hereafter AIA~193) was found to be $\sim$21~h. Convergence and cancellation of the CBP bipoles typically take place, both before and during the eruptions. The CBP eruptions unfold with the expulsion of chromospheric material either as an elongated filamentary structure (mini-filament) or as a volume of cool material (cool plasma cloud). This is usually accompanied by the ejection of the CBP or/and higher overlying hot loops. Occasionally coronal waves are also observed. Micro-brightenings called micro-flares are detected in all eruptions and are always associated with the polarity inversion line (PIL) of the bipoles related to the eruptions. 
The nature of the micro-flares is still to be determined. Mini coronal mass ejections (mini-CMEs) occur in 11 out of the 21 CBP eruptions. Dimmings linked to the propagating CMEs are seen as both `dark' cool plasma and areas of decreased coronal emission resulting from a plasma density depletion. This indicates the possibility that mini-CMEs represent a characteristic part of the general CBP lifecycle, and that it is a natural stage in the evolution of CBPs. 

In Paper~II, the non-potential time dependent structure of the magnetic field of the CBPs   from Paper I was investigated at the spatial locations of the eruptions. This investigation also considered the nature of the overlying coronal magnetic field above each CBP.    To carry out the investigation a NLFFF relaxation approach, based on  a time series of Helioseismic and Magnetic Imager (HMI) line-of-sight magnetograms was used to produce a continuous time sequence of NLFFFs for the CBPs.  In each case the initial condition was taken to be a potential field extrapolation based on a magnetogram taken before the eruption time. This initial field was then evolved in time in response to the observed changes in the magnetic field distribution at the photosphere. The local and global magnetic field structures from the time series of NLFFF fields  were then analysed in the vicinity of the eruption sites at the approximate times of the eruptions. The analysis shows that many of the CBP eruptions reported in \citet{2018A&A...619A..55M} contain  magnetic flux ropes  at the spatial location of the eruptions. The presence of flux ropes at these locations, provides, in many cases a direct link between the magnetic field structure, their eruption and the observation of mini-CMEs. It is found that all repetitive eruptions are homologous. The NLFFF simulations show that twisted magnetic field structures are created at the locations hosting eruptions in CBPs, where the flux ropes are produced by the footpoint motions occurring in the photospheric magnetic field observations. Despite this advance in our knowledge of mini-solar eruptions, the true nature of the micro-flares remains unclear. 

\begin{figure}[!ht]
\includegraphics[scale=0.7]{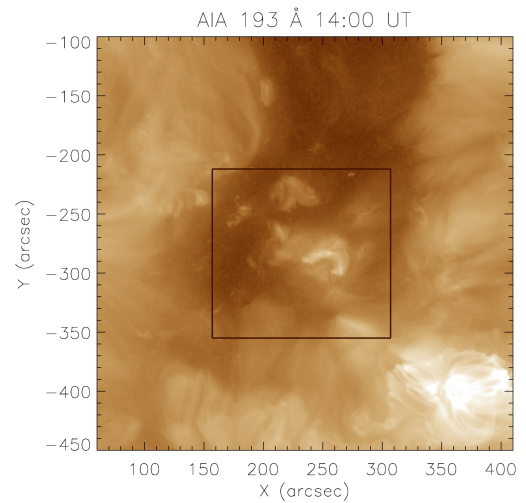}
\caption{AIA 193 image showing an equatorial coronal hole. The coronal bright point that is the subject of the present study is located at the centre of the field-of-view. Overplotted with black solid line is the field-of-view shown in Fig.~\ref{fig2}.}
\label{fig1}
\end{figure}

The present study investigates a single case of a coronal hole CBP eruption  (Fig.~\ref{fig1}) that was caused by the destabilization and eruption of a mini-filament. The eruption resulted in the formation of a jet seen in X-rays. This CBP was selected from a collection of several eruptive CBPs identified in simultaneous X-ray and EUV data. In addition, the CBP eruption was recorded in H$\alpha$ observations which adds crucial information on the response of the solar chromosphere to the flaring activity of the CBP. These observations, combined with NLFFF modelling permit 
us to investigate in full detail the connectivity between the solar chromosphere and corona, and the build-up to an eruptive state of a simple small-scale magnetic loop system in the solar atmosphere.  It is important to note that the observations may be used to understand both the pre-eruptive, eruptive and post-eruptive structures. In contrast the NLFFF modelling can only be used to
understand the pre-eruptive and build-up to eruption properties of the magnetic field.
The uniqueness of the chosen CBP relates to its size which exceeds the typical upper limit of CBP sizes of $\sim$60\arcsec, covering a solar-disk projected area with a diameter of more than $\sim$100\arcsec. This large CBP gives a unique opportunity to observe and model fine details of the CBP eruption that are often affected by the spatial resolution. 
In addition, we are able to identify the physical nature of the micro-flares, filament eruptions,  X-ray/EUV jet formation and more importantly  the connectivity and thus the energy transport between the corona and chromosphere during the small-scale solar eruption. 

The paper is organized as follows.  Section~\ref{obs} provides detailed information on the analysed observational material. The observational results are given in Section~\ref{res.obs}, and the outcome from the modelling work is presented in Section~\ref{res.model}. The obtained results are discussed in Section~\ref{disc}. The inferred conclusions of the present investigation are given in Section~\ref{concl}.

\section{Observational material}
\label{obs}

The event occurred on 2013 October 12. To study the response of the solar chromosphere to the 
energetic (micro-) flaring event we used data from the GONG H$\alpha$ observational program. 
The data represent images taken with a Daystar H$\alpha$ filter with a bandwidth of 0.67~\AA\ targeting
a wavelength of 6562.8~\AA, but shifted to the red. The H$\alpha$ data cover the time from 15:18~UT until 16:59~UT, with a cadence of 60~s. Images from 11:54~UT until 14:54~UT are also available at 1~h cadence and were used to determine the formation time of the mini-filament (hereafter MF or simply filament). 
The X-ray observations analysed for this study were obtained with the X-Ray Telescope \citep[XRT;][]{2007SoPh..243...63G} on board the Hinode satellite at 1\arcsec\ plate scale and a cadence of 5.7~min using 
the Al\_poly filter. We also used data from  AIA \citep{2012SoPh..275...17L} on board SDO \citep{2012SoPh..275....3P}, which consists of seven Extreme Ultra-Violet (EUV) and three Ultra-Violet (UV) channels providing an unprecedented view of the solar corona with an average cadence of $\sim$12~s. For our analysis we employed images taken at 1~min cadence in the EUV 304, 171, 211, 193 and 94~\AA\ channels (hereafter e.g. AIA~304). Line-of-sight magnetograms taken by  HMI \citep{2012SoPh..275..207S} on board SDO covering the time period between 07:00 and 17:53~UT at a time cadence of 10~min were used in the NLFFF modelling. The HMI magnetograms and AIA images were co-aligned by using the UV AIA~1600~\AA\ channel that was consequently aligned with the AIA EUV channels. All data were de-rotated to 07:00~UT on 2013 October 12.

\section{Results}
\label{res.obs}

The CBP under investigation in the present study, had an exceptionally long lifetime of more than 9 
days. Such a long lifetime is however not surprising given its large size. It has long been established that the lifetime of CBPs is approximately proportional to their maximum size \citep{1974ApJ...189L..93G}. Figure~\ref{fig1} shows the CBP (in the centre of the black-lined square) located at the southern end of an equatorial coronal hole. A bundle of bright loops are seen surrounded by diffuse emission. The precise determination of the lifespan of this CBP is hard to make as both the formation and end locations are close to the limb 
(east $\sim$ formation and west $\sim$ end).  
Generally, when CBPs are the result of flux emergence at both their `birth' and `death', 
the CBPs tend to be rather small (diameter 5\arcsec\ or lower) \citep{2018A&A...619A..55M}.

\begin{figure*}[!ht]
\hspace{-1.cm}
\includegraphics[scale=0.5]{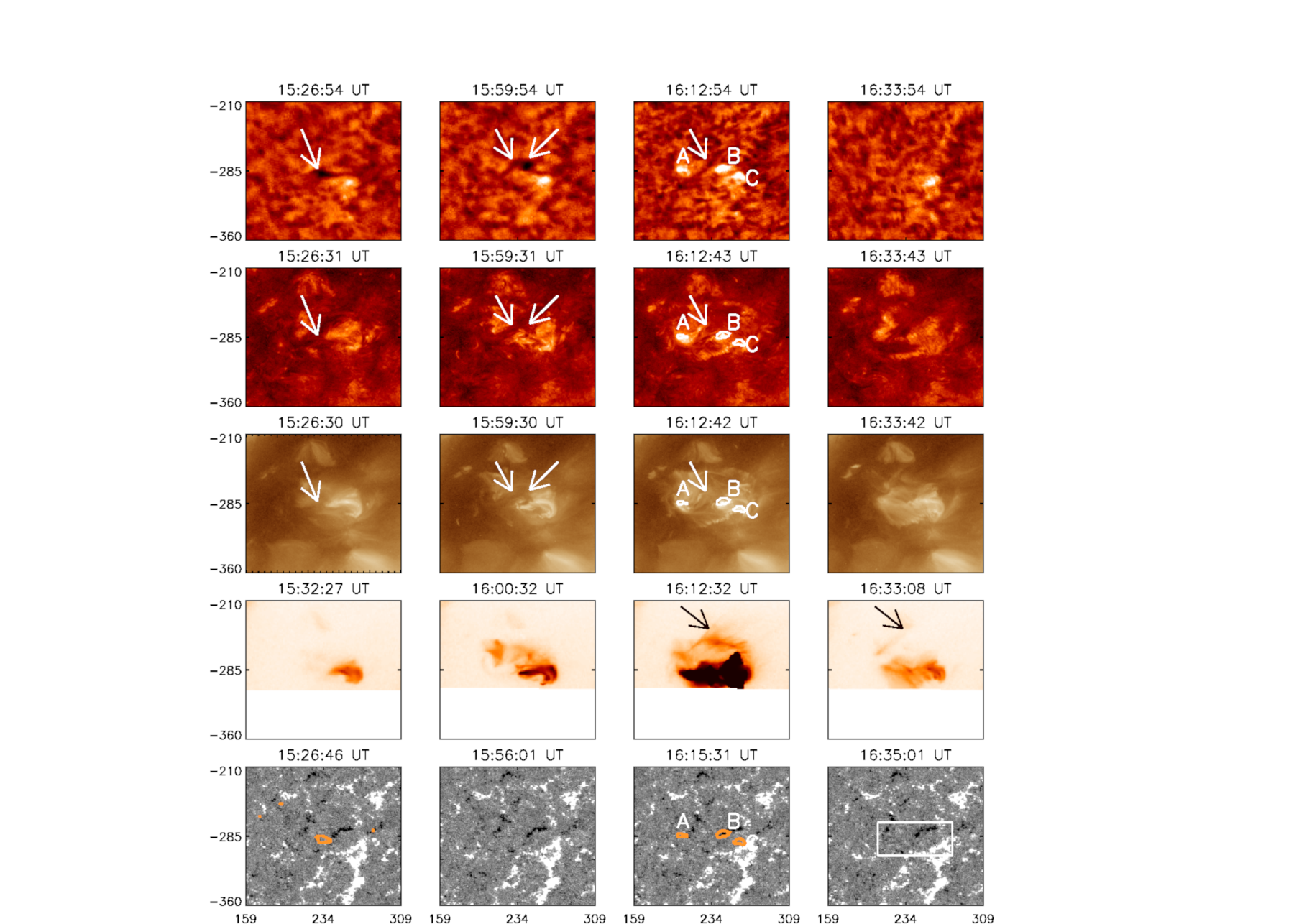}
\caption{From top to bottom rows: GONG H$\alpha$, AIA~304, AIA~193, XRT and HMI images showing co-observations of the pre-eruption (first column), eruption (second) and post eruption (columns three and four) phase of the CBP. White arrows point at the erupting filament. The postflare loops can be seen in the third and forth columns of AIA~193 and XRT images. The black arrows on the XRT images point at the X-ray jet (the images are shown with reversed (negative) colours). The HMI magnetograms are scaled from $-$50 to 50~G. The orange contour on the magnetograms (first panel) outlines the largest part of the mini-filament. The orange contours (third panel) outline the location of the micro-flare kernels noted with  A, B, and C.}
\label{fig2}
\end{figure*}

\subsection{Observational analysis and results}
\label{sec.obs}

The eruption investigated here is one of a series of six eruptions that originated from the studied CBP. We focus our analysis on the eruption on October 12 as this event is recorded by Hinode/XRT revealing the formation of a collimated outflow, i.e. a jet. We note that XRT can take observations during 
both specially designed and time-allocated campaigns, with for instance a limited field-of-view (FOV). Data from the GONG H$\alpha$ observing survey program are also available (see Section~\ref{obs}) which gives a rare chance to study the chromospheric response to a small-scale eruptive phenomenon. 
The event investigated here was the largest of all 
six of the CBPs eruptions. None of these events resulted in the disappearance of the CBP. The event was identified during a dedicated search for X-ray jets occurring in equatorial holes from within  the XRT archive. As mentioned above this jet was selected for further analysis due to the CBP size, although a large number of equatorial-region jets were also found. The eruptive phenomenon took place while the CBP was located at solar heliographic coordinates xcen = 320\arcsec\ and ycen=-270\arcsec, where xcen and ycen are the approximated CBP centre coordinates on 2013 October 12. 

\begin{figure*}[!h]
\centering
\vspace{-2.5cm}
\includegraphics[scale=1.2]{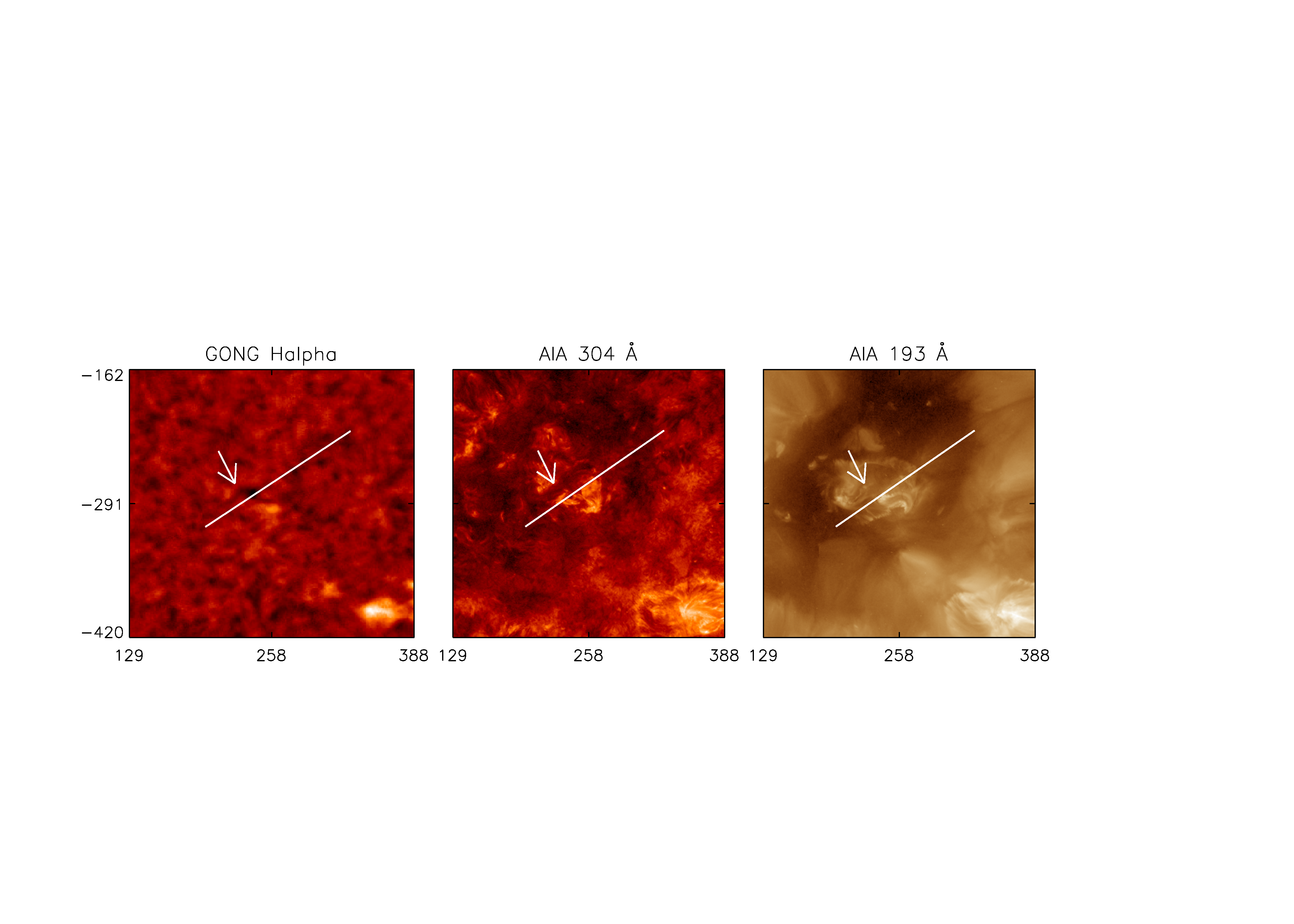}
\vspace{-4cm}
\caption{From left to right: GONG H$\alpha$, AIA~304, and AIA~193 images showing the CBP with the erupting filament visible in absorption in the all three channels. The white solid line is the location from which the time-slice image in H$\alpha$, several AIA channels and XRT are shown in Fig.~\ref{fig4}. The images are taken at 15:58~UT. White arrows point at the erupting filament.}
\label{fig3}
\end{figure*}

The CBP eruption follows the general scenario of evolution already seen during Quiet Sun (QS) CBP eruptions \citep[for details see][and the following paragraphs]{2018A&A...619A..55M}. Because of the CBP's large size and the availability of data that cover a wide range of temperatures (including H$\alpha$ -- chromosphere and X-ray -- high temperature corona), we were able to observe, model, and thus understand important details of this eruptive phenomenon. As mentioned already, the CBP is located in a coronal hole, a region dominated by low emission and an open magnetic field. As expected \citep[e.g.][]{2011A&A...529A..21K}, rather than evolving into a  mini-CME, 
as usually happens in closed coronal magnetic-field topologies in QS regions  \citep[for details see][]{2009A&A...495..319I,2010A&A...517L...7I, 2018A&A...619A..55M}, a collimated flow, namely a jet, is seen 
in both EUV and X-ray emission \citep[e.g.][and the references therein]{2016SSRv..201....1R}. The jet-like eruption was associated with the formation, destabilization and eruption of a MF recorded in both the EUV and H$\alpha$ observations. Below we present and discuss details on the pre-eruption, eruption and post-eruption phases  as deduced from the observations.

\subsubsection{Pre-eruption phase}
Before the eruption a MF is seen in the H$\alpha$ images at the location of the CBP as early as 12:54~UT. Table~\ref{table1} outlines the timeline of the series of events described below. An earlier H$\alpha$ image taken at 11:54~UT does not show the MF and therefore, the MF formed within the time interval of 11:54~UT and 12:54~UT. At its earliest detection in the H$\alpha$ images the MF cannot be identified in the AIA~304 and 193 images. In the AIA~304 channel the MF cannot be separated from other `dark' structures that could be either cool material \citep[see][for details]{2018A&A...619A..55M} or simply the result of a lack of emission at transition region or coronal temperatures. In the AIA~193 channel the filament is possibly obscured by the overlying coronal loops of the CBP. \citet{2018A&A...619A..55M} has already pointed out that MFs are often not visible until the time of their eruption. This may be because they lie very low in the solar atmosphere, i.e. below the CBP loops that have an average height of 6\,500~km \citep{2019LRSP...16....2M}, or they form only shortly (an hour or so) before their eruption. The investigation of \cite{1986NASCP2442..369H} reports that the time from MFs' formation to their eruption is on average 70~min.  At the present time it is unknown whether their end of life always results in an eruption. 

The MF is most clearly seen in the first H$\alpha$ image of the 60~s cadence sequence at 15:18~UT (see Fig.~\ref{fig2}, where an image from 15:26~UT is shown as it was the first high
quality image from the time series). At this time it is still difficult  to distinguish the MF among the other dark structures seen in the AIA~304 and 193 images. However, from knowing the MF location in the H$\alpha$ image, we are able to also identify it in the EUV images (see the arrows in the first column of Fig.~\ref{fig2}). This illustrates that chromospheric data (e.g. H$\alpha$) are essential in studying any eruptive solar 
phenomenon, as the physical processes involved in their formation and evolution leave important footprints throughout the whole solar atmosphere. The two arrows on the images in the second-column in 
Fig.~\ref{fig2}, taken at 15:59~UT, point out the MF just as it starts to rise at approximately 3--4~h after its formation. The footpoint separation of the MF was estimated 
to be 60$\pm$5\arcsec\ using the image at 15:59~UT, where the MF is clearly visible along its whole length.
 
\begin{table*}[h!]
\caption{Timeline of the CBP--MF eruption.}
\label{table1}
\centering
\begin{tabular}{  c  l  }
\hline
Time  (UT) & Event\\
\hline
11:54 -- 12:54 & MF formed in this time interval\\
15:35 -- 15:57 & MF starts to slowly rise, at a few km/s\\
15:57 -- 16:05 & Fast eruption of MF at 30$\pm$5 km/s\\
15:57          & Micro-flare kernel B appears\\
16:01          & First coronal brightening in the EUV channels\\
16:02 \& 16:04 & Jet seen in EUV \& X-rays\\
16:07          & The micro-flare ribbon appears\\
16:10          & Some of the MF is ejected along the open field lines\\
               & A dimming appears in AIA~304\\
16:20          & The ribbon fades away in H$\alpha$\\
16:40          & The dimming fades away\\
               & The ribbon disappears in EUV\\
               & X-ray emission fades away\\
\hline
\end{tabular}
\end{table*}
As expected, the MF lies along the polarity inversion line that separates the bipoles forming the CBP. 
The contour on the HMI magnetogram in Fig.~\ref{fig2} (first panel) outlines the darkest feature in the H$\alpha$ image taken at 15:26~UT. This is the densest and widest part of the MF. It should be noted that the position of the contour on the HMI magnetogram (at photospheric heights) is affected by projection effects as the CBP is not located at disk centre. This can explain why the MF location is not precisely between the positive and negative polarities. The modelling of the coronal magnetic field and 
that of the MF flux rope also shows that the position and the flux rope connectivities are more complex than what can be deduced by simply overlying the H$\alpha$ position with that of the polarities on the photospheric magnetogram (see Section~\ref{res.model} for the model details).

The observed unsigned total photospheric magnetic flux (see Fig.\ref{app1}) shows a continuous steady decrease of $\sim$15\% for the time period between 08:00 and 18:00~UT (estimated from a boxed region at the location of the CBP). The two major opposite polarities show a very small convergence but they remain at a large distance from one another during the analysed time period. Therefore, flux cancellation between these opposite 
magnetic polarities can be excluded as the main cause for the flux rope formation and the subsequent MF eruption. At the north-eastern edge of the negative polarity, a small-scale flux emergence event   
takes place, followed by flux divergence and then convergence and cancellation. This new flux does not appear to have a marked influence on the general trend of decreasing flux in the region. Observed 
motions and magnetic flux cancellation of the small-scale flux concentrations may have been the mechanism 
for the build up of the flux rope (see  Section~\ref{res.model} for further details).

\begin{figure*}[!ht]
\centering
\includegraphics[scale=0.75]{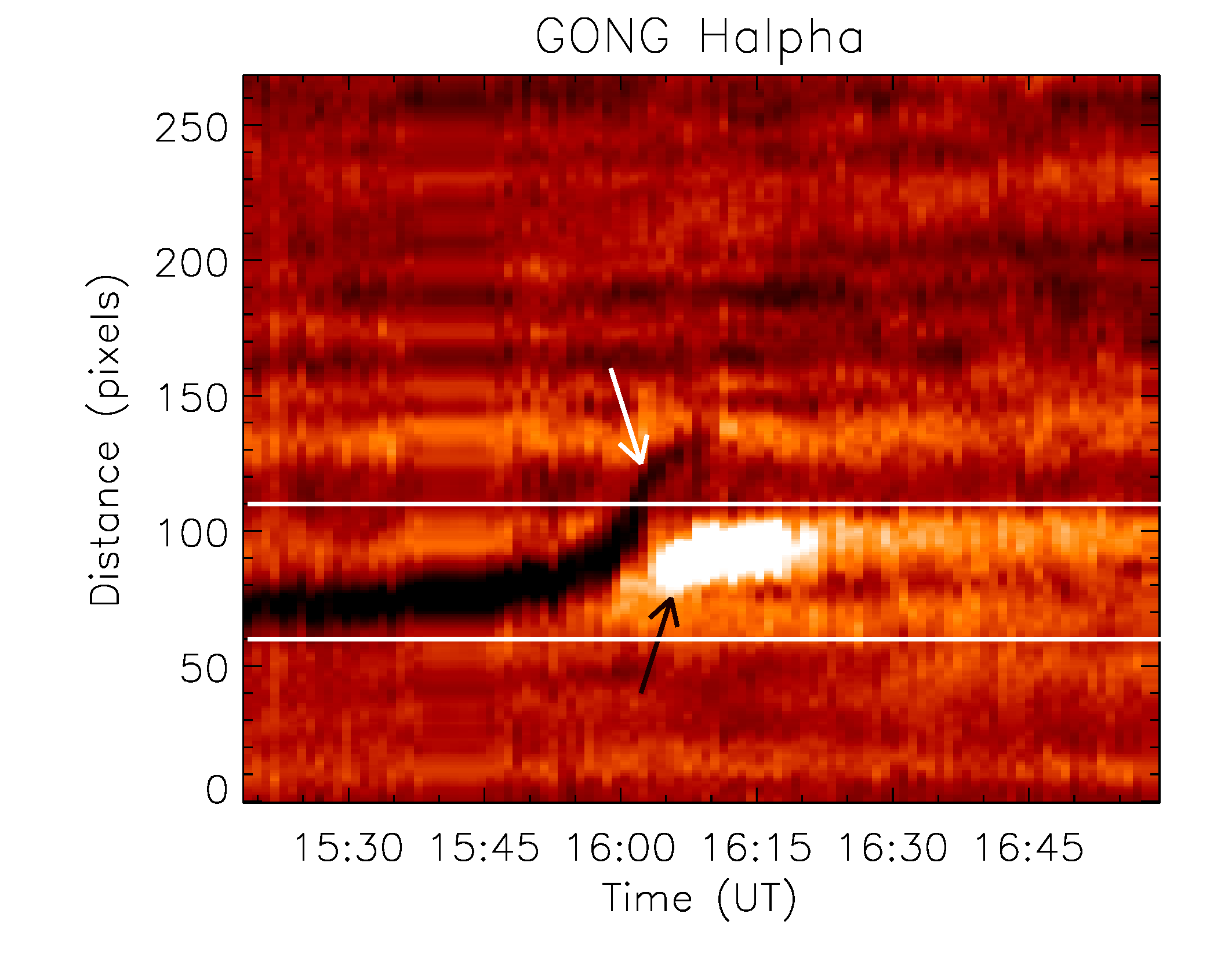}
\includegraphics[scale=0.75]{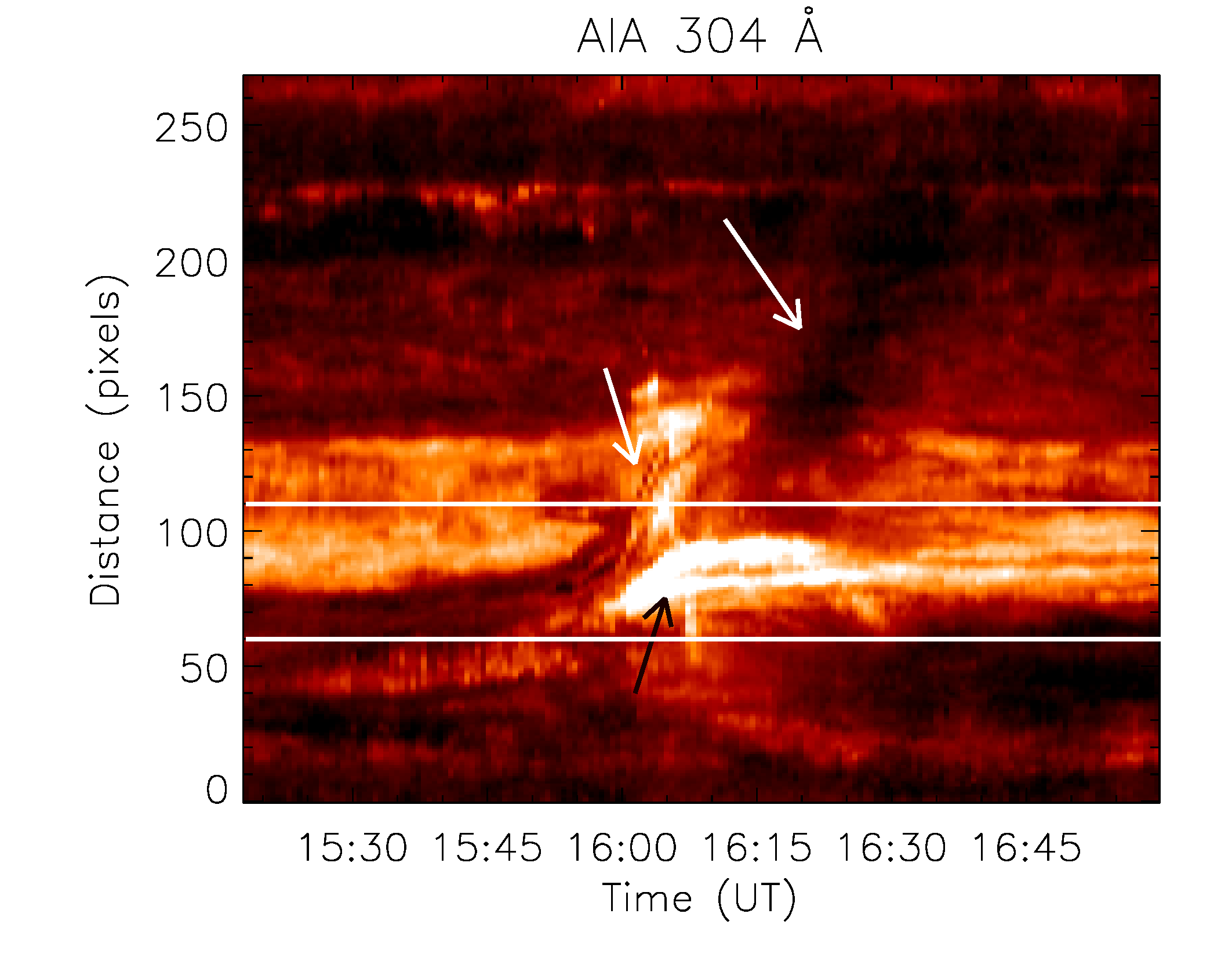}\\

\includegraphics[scale=0.75]{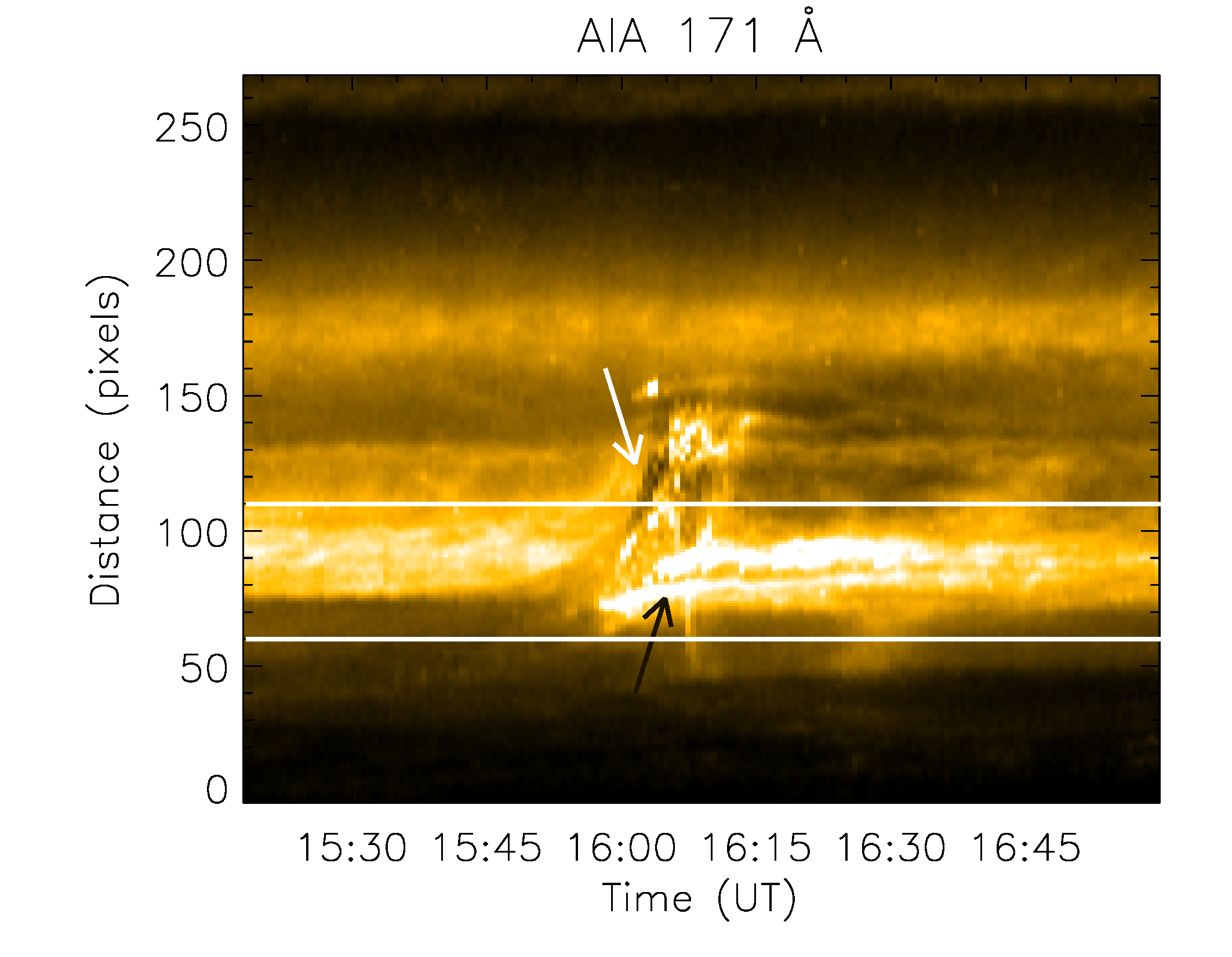}
\includegraphics[scale=0.75]{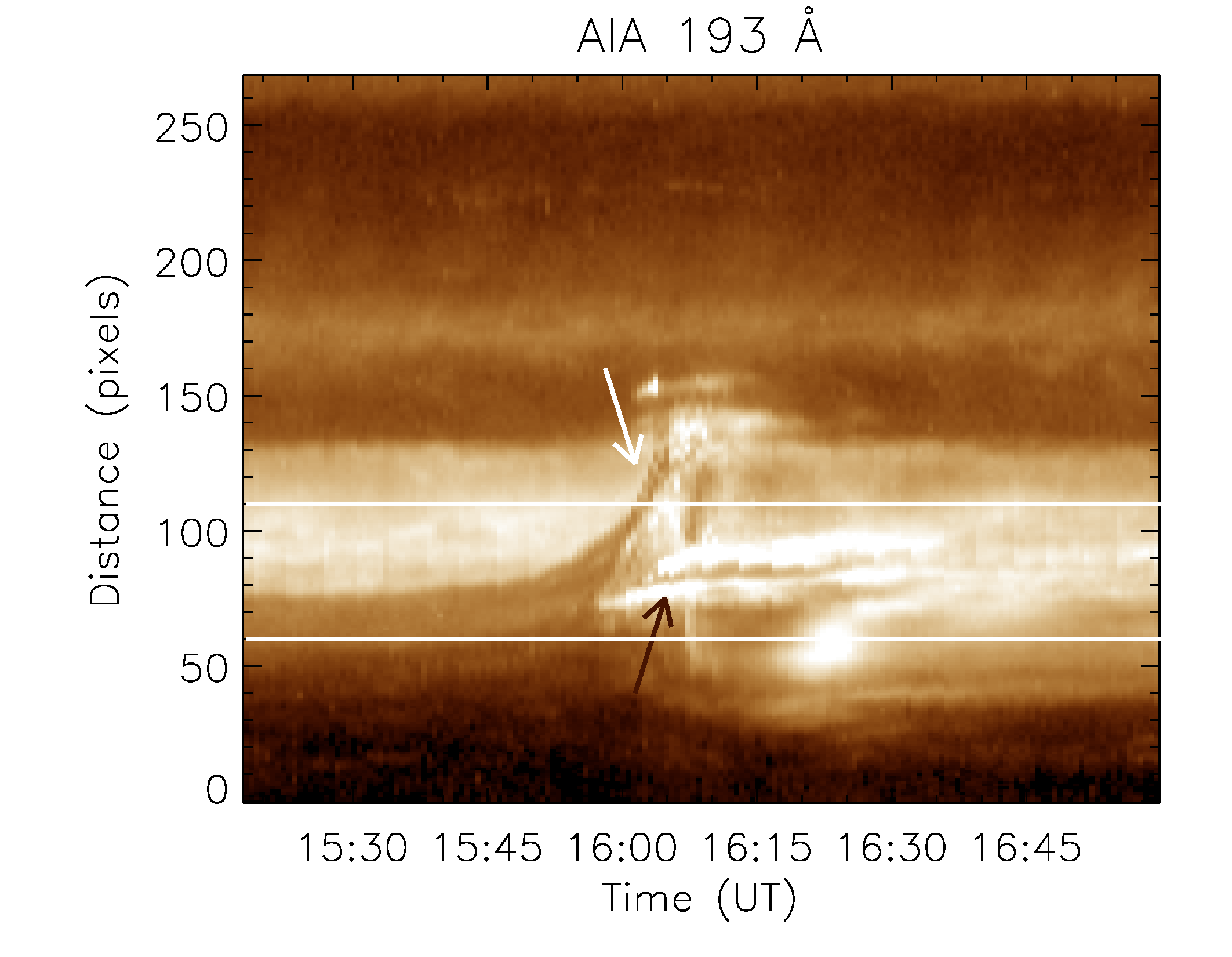}\\

\includegraphics[scale=0.75]{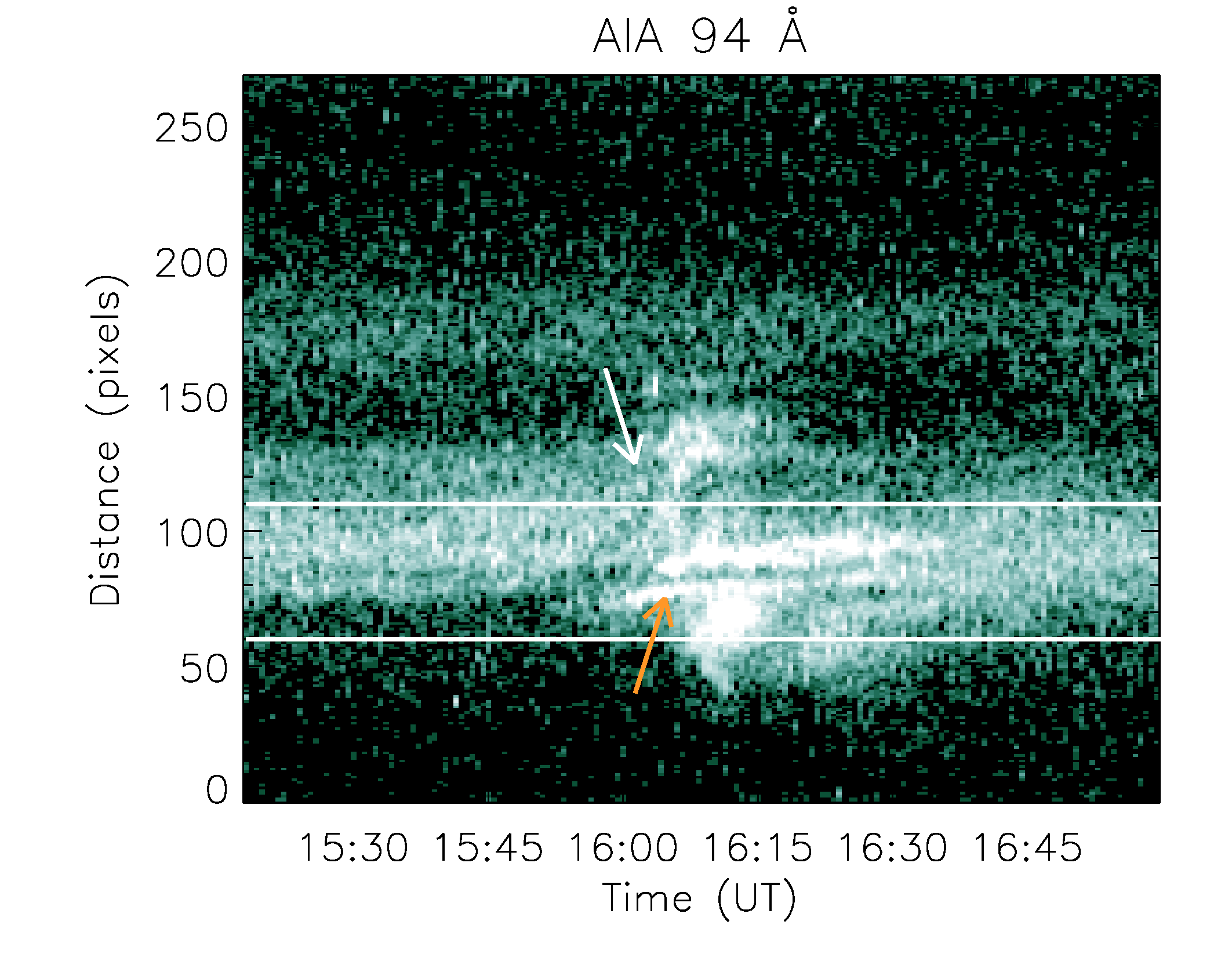}
\includegraphics[scale=0.75]{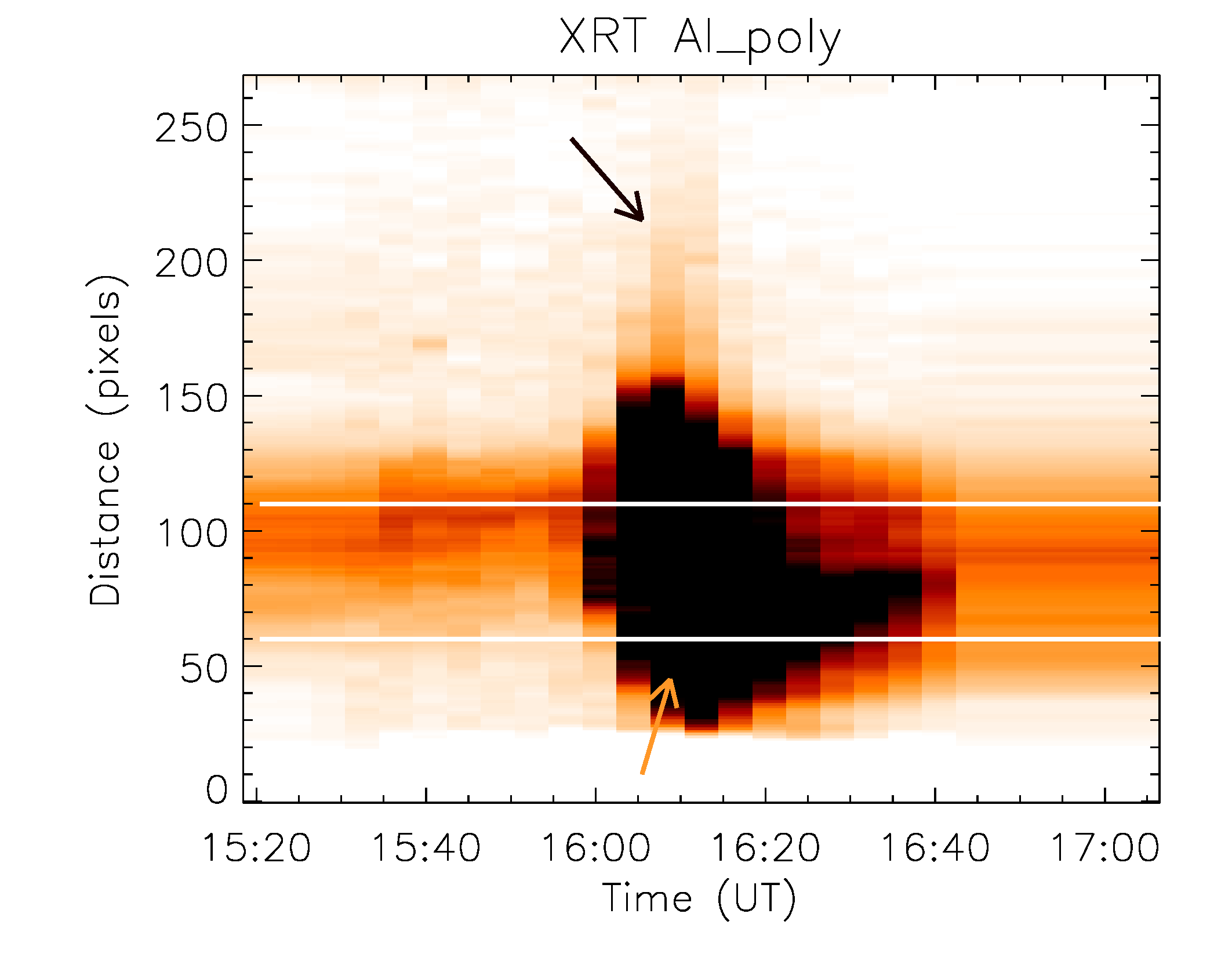}
\caption{Time-slice images in H$\alpha$, AIA~304, 171, 194 and 94 channels, and X-rays produced from the location shown in Fig.~\ref{fig3}. White arrows point at the erupting filament. The white upper arrow on the AIA~304 image points at the dimming. The black arrow on the XRT image points at the X-ray jet. The white/orange bottom arrows indicate the micro-flare. The two horizontal white lines outline the region from which the lightcurves in Fig.~\ref{fig5} were produced. }
\label{fig4}
\end{figure*}

\subsubsection{Eruption phase}
The eruption phase of the MF follows that typically found for active-region or quiescent filaments \citep[][and the references therein]{2014LRSP...11....1P}. The MF started to ascend slowly at $\sim$15:35~UT with a speed of a few kilometers per second where this lasted until approximately 15:57~UT. The slow rise was then followed by a fast lift-off of the MF. Figure~\ref{fig3} shows GONG H$\alpha$, AIA~304, and AIA~193 images with an overplotted line that passes through the filament. The emission along this line was used to create the time-slice images shown in Fig.~\ref{fig4}  that illustrate the temporal and spatial evolution of the MF eruption, as well as one of the micro-flare kernels (kernel B, see the following paragraph for more details). Flare kernels and ribbons are one of the main observational components of solar flares as they represent the location of chromospheric heating.  From the time-slice H$\alpha$ image we estimated an eruption speed along the slice of $\sim$30$\pm$5~\kms\ (the rise of the MF between 15:57 and 16:05~UT). This speed should not be considered purely as an upward motion. The provided animation 
(Fig.~\ref{app2}) demonstrates that some of the MF plasma that was located in the magnetic flux rope lifts up, swirls and follows the open magnetic field lines of the already formed coronal jet escaping into the upper corona. The darkening/dimming strip seen in the AIA~304 time-slice image after 16:10~UT illustrates this cool material propagation which appears to evolve at the same speed of $\sim$30$\pm$5~\kms. 

At 16:01~UT the AIA~304, 193, 171 and 94 images in Fig.~\ref{fig4} reveal the first coronal brightening above the filament that could be (but not solely) related to magnetic reconnection associated with the rising MF  magnetic field. Increased emission accompanying the whole rising filament can also be seen in the AIA~304, 193, 171 and 94 images of Fig.~\ref{fig4}. 

A signature of a micro-flaring event is first and best seen in the AIA~171 channel at 15:57~UT.
The event indicates that chromospheric heating occurs at the pre-eruption location of the  MF at the very start of the fast lift-off phase. Three distinctive bright kernels, A, B and C as indicated in Fig.~\ref{fig2} are observed in the H$\alpha$ images.  All three are a part of the micro-flare ribbon that forms at a later time. The kernels are well observed in the EUV channels, however without the information from the H$\alpha$ images, it would have been hard to identified them as flare kernels given the complexity of the emission seen in the EUV channels during the eruption. As mentioned above, the path of the cut in Fig.~\ref{fig3} slices through the rising MF and one of the flare kernels, kernel B. It is well known from flare studies that flare kernels result from rapid intense heating that occurs during the rising phase of solar flares. They are believed to result from non-thermal particles that are released during magnetic reconnection in the corona \citep[e.g.,][]{1985ApJ...289..414F,2013ApJ...766..127Y}. This reconnection occurs between the ascending flux rope of the MF and the overlying loop structures \citep{2007ASPC..368..365H, 2011SSRv..159...19F,2012ASPC..456..183F}. No observational signature is found for reconnection beneath the filament. Section~\ref{res.model} discusses the 
relationship of the kernels to that of the coronal magnetic field configuration.

\begin{figure*}[!ht]
\hspace*{1.cm}
\includegraphics[scale=0.8]{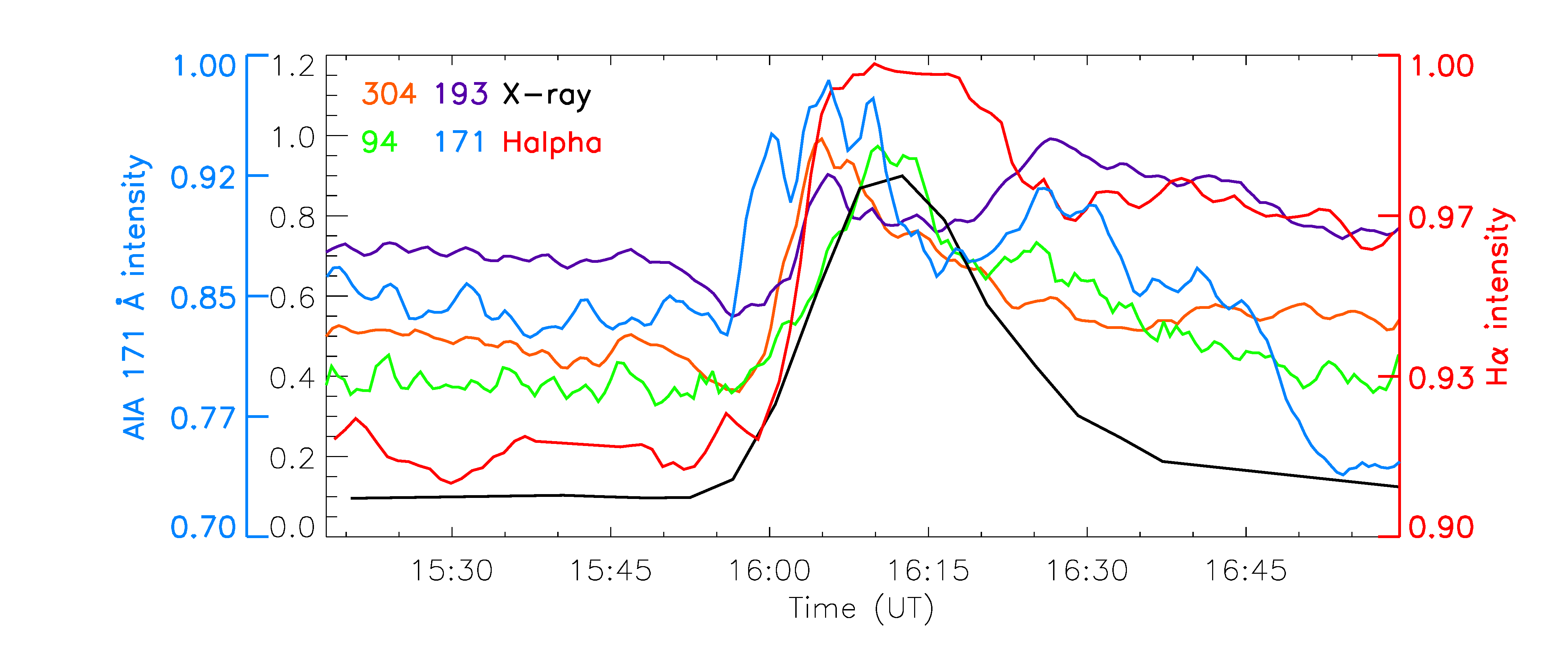}
\caption{Normalized (to the maximum intensity value) lightcurves in H$\alpha$ -- red line with corresponding right Y axis, AIA~304 -- orange, AIA~193 -- purple, AIA~94 -- green, and X-rays (black, black left Y axis), as well as AIA~171 (blue, blue left Y axis), produced from the location outlined with two horizontal white lines in Fig.~\ref{fig3}.}
\label{fig5}
\end{figure*}

The lightcurves in Fig.~\ref{fig5} show the  variation of the intensity in the different wavelength channels where the intensity temporal evolution is determined from the area between the two horizontal lines shown in Fig.~\ref{fig4}.  In Fig.~\ref{fig2} and the animation (Fig.~\ref{app2}), one can see that the time-slice image reveals the evolution of the rising MF as well as the micro-flare kernel B as described above. The earlier response of the AIA~171 channel compared to the other EUV channels (94, 193 and 304~\AA), can possibly be explained by the fact that this channel includes low temperature emission 
from the O~{\sc v} 172.2~\AA\ and O~{\sc vi} 173.0~\AA\ lines. Thus the MF is not seen strongly in absorption, but only with slightly reduced emission. Therefore, the response in this channel is the true starting time of the kernel appearance. The early response in the EUV channels compared to the X-ray emission is consistent with an energy transportation process dominated by non-thermal particle beams generated during the reconnection process. The later response in H$\alpha$ comes from the obscuring of 
kernel B by the MF along the line of sight (see the animation in Fig.~\ref{app2}). The increase 
of emission in X-rays is delayed by 5--6 min. It is dominated by the emission from the coronal loops
heated to X-ray temperatures where heat conduction is the dominant energy transport mechanism. Soft X-ray emission may also have been emitted from the heated chromosphere, but the observations are dominated by the bright coronal structures that obscure the kernels and thus they cannot be seen. The lightcurves reveal a fast rising micro-flare phase followed by a gradual phase as typically observed in solar flares \citep{2011SSRv..159...19F}.

The evacuation of hot material in X-rays and EUV is first seen as a collimated flow (a jet) along open magnetic field lines, starting around 16:02~UT in AIA~304 or 16:04~UT in the XRT images (the XRT cadence is 4~min). The jet evolution can be followed in the provided animation that shows quasi-cotemporal AIA~304, AIA~193 and XRT  (Fig.~\ref{app2}) images. This is followed at 16:10~UT by a cloud of cool plasma (partially erupting MF) seen in absorption in the AIA~304 images as mentioned above. The jet is not visible in the AIA~193 and 171 images, because of the strong background emission in these channels as the jet propagates above the QS.

\subsubsection{Post-eruption phase}

The postflare loops gradually become brighter and are most visible after 16:20~UT (see the right column of Fig.~\ref{fig3}), $\sim$20~min after the micro-flare takes place. The lightcurves in Fig.~\ref{fig5} display a typical gradual phase with a slow intensity decrease of the fading (cooling) post-flare loops and ribbons. The micro-flare ribbon as well as the flare kernels fade away first in the H$\alpha$ images (around 16:20~UT) lasting in the EUV until at least 16:45~UT.

\subsection{Modeling analysis and results}
\label{res.model}

\begin{figure*}[!ht]
\centering
\includegraphics[scale=0.17]{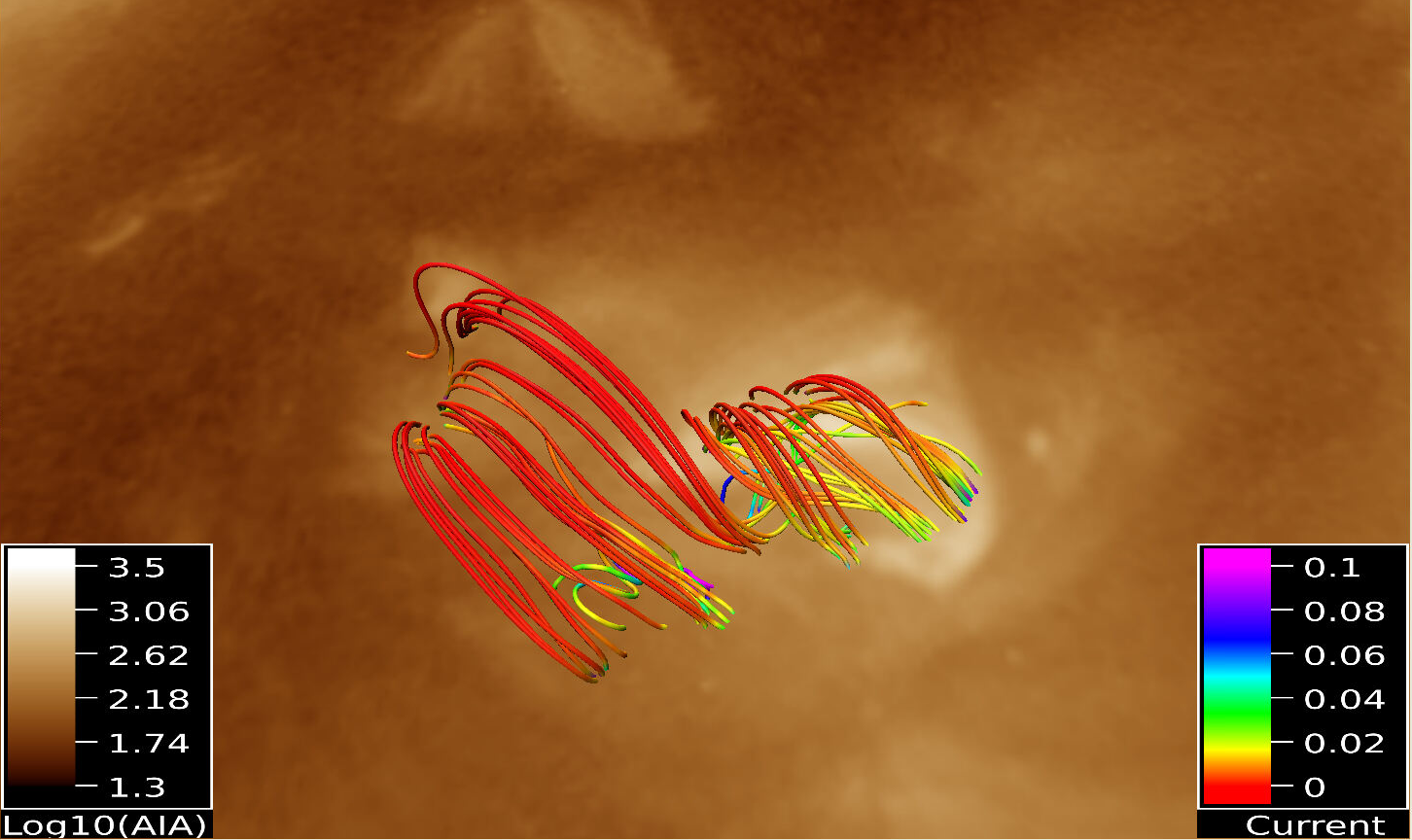} 
\includegraphics[scale=0.17]{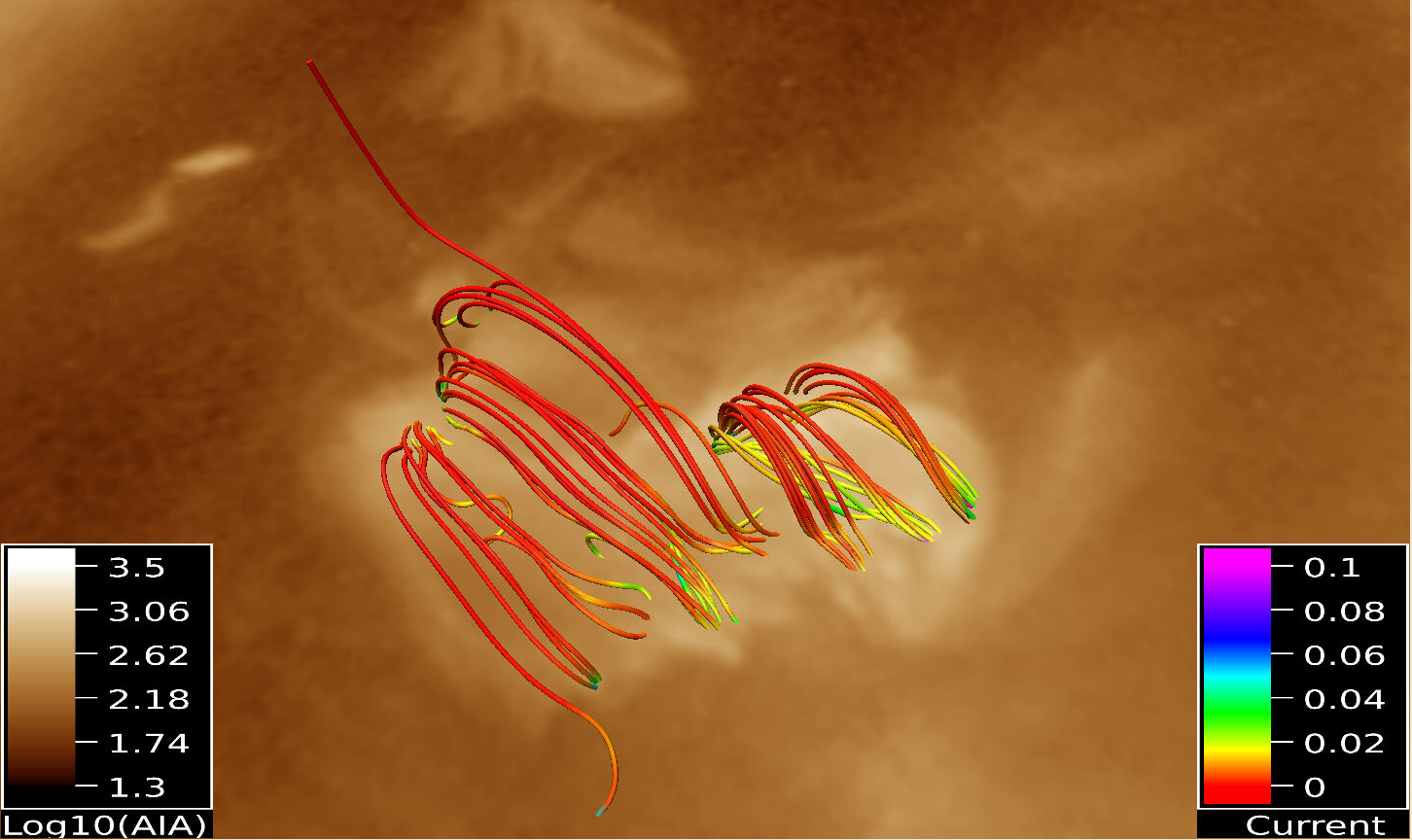}\\ 
\includegraphics[scale=0.17]{f6c_38287.png} 
\includegraphics[scale=0.17]{f6d_38287.png} 
\caption { Left column: AIA~193 and HMI taken at 15:26~UT, with overplotted the NLFFF  structure. Right column: the same as for the left column but at 16:35~UT. The left and right columns correspond respectively to the time  prior the eruption and post-eruption phases. }

\label{fig6}
\end{figure*}

\begin{figure*}[!ht]
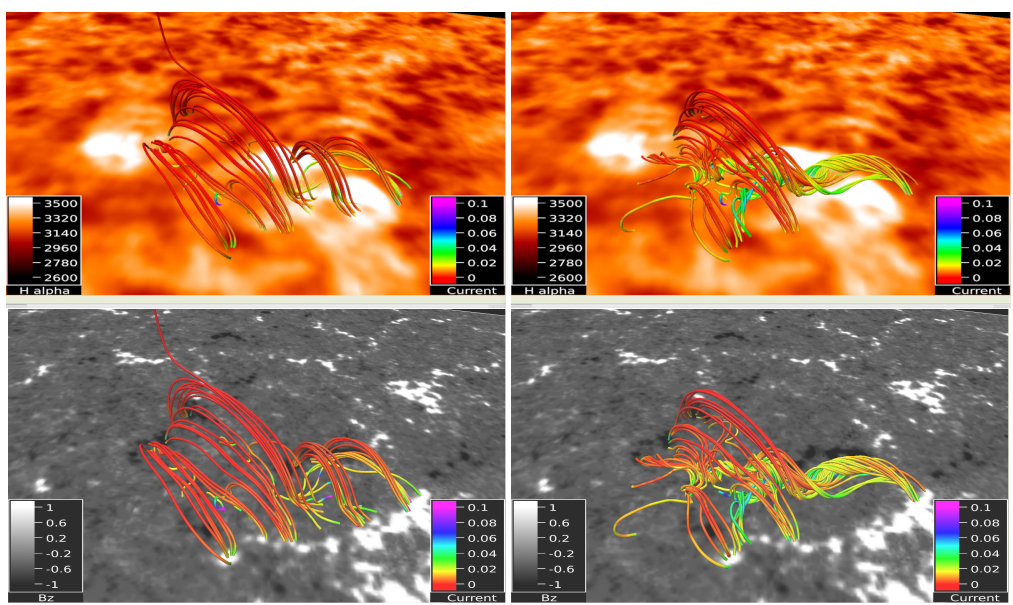

\centering
\hspace*{-1cm}
\includegraphics[scale=0.31]{f7a_38287.png} 
\includegraphics[scale=0.31]{f7b_38287.png}\\ 
\hspace*{-1cm}
\includegraphics[scale=0.17]{f7c_38287.png}
\includegraphics[scale=0.17]{f7d_38287.png}
\caption{ Top row: H$\alpha$ images taken at 16:13~UT during the eruption phase with overplotted overlying coronal field  (left panel) and the magnetic flux skeleton that holds the filament (right panel). Bottom row: HMI magnetograms taken at 16:15~UT with the same overplotted magnetic field lines.}
\label{fig7}
\end{figure*}

The NLFFF modelling, based on the time series of HMI data (covering the time from 07:00~UT to 17:53~UT on 12 October), follows the procedure discussed in \citet{2019A&A...623A..78G}. Here we provide a brief introduction to the modelling approach where full details along with a justification for using this approach is given in Appendix A. It is important to note that this modelling approach may only be used to study the coronal field of the CBP and MF during the build-up to the eruption to understand the nature and evolution of the pre-eruptive field. It cannot be used to simulate the eruption itself. The initial 3D condition of the magnetic field is chosen to be a potential magnetic field derived from the first HMI magnetogram in the time series. 
This 3D magnetic field is then evolved forward in time using the observed changes in the photospheric magnetic flux found in the time series of the HMI data. These changes inject electric currents and non-potentiality into the 3D coronal magnetic field. After each change in the HMI magnetograms is applied, the 3D configuration is relaxed to a new NLFFF solution, before the boundary conditions are once again changed and the process repeated. This allows the 3D magnetic field to evolve through a series of NLFFF configurations based on the applied boundary motions. Such an evolving magnetic field configuration enables for the buildup of free magnetic energy which may eventually be converted into eruptive events if/when the magnetic structure becomes sufficiently stressed. This semi dynamical evolution of the NLFFF cannot be reproduced by deriving a time series of static NLFFF models from a series of independent well defined vector field boundary conditions. The NLFFF modelling approach described above relies only on normal component magnetograms which is important as HMI vector magnetograms can only be used in regions with strong magnetic fields due to the low signal-to-noise ratio of the transverse component of the field in the quiet Sun \citep{2014SoPh..289.3483H,2009SoPh..260...83L}. As the noise level of the HMI transverse component is 100~G \citep[e.g.][]{2013A&A...550A..14T} and the CBP transverse component is far below this value, HMI vector magnetograms cannot be used in the present study. 

Figure~\ref{fig6} shows how the coronal loop structure forming the CBP in the NLFFF modelling changes over a 69~min time period around the time of the  observed eruption. The field lines are plotted on top of a background image taken in the AIA~193 channel (top row) and HMI magnetograms (bottom row), respectively. The magnetic field lines are traced from fixed positions in time close to the photospheric surface.  
Each magnetic field line is colour coded based on the strength of the electric current along the field line, where the strength of the current is given in the panel on the lower right of each image. The frames show two sets of loop systems that connect from the large  positive polarity to two separate negative flux concentrations. These two large loop systems show only small changes over the 40 min time frame, even though this time period covers the time of the eruption, where significant changes are seen in the coronal AIA~193 images. This is not surprising as a recent model  of X-ray jets from CBPs by  \citet{2018ApJ...864..165W}  has shown similar results. 
 The observations show that after the postflare loop formation,  the observed loops relax back to their earlier configuration.
This is supported by the findings of the NLFFF model where the pre-eruption arcades are close to potential. 

During the pre-eruption phase, the smaller and more compact magnetic field  arcade in the left panel of Fig.~\ref{fig6} clearly outlines the location of the enhanced emission in the AIA~193 image related to the CBP. The larger arcade is associated with  a rather hazy enhanced coronal emission  where no clear structuring is visible. In the  observed postflare/eruption phase the plasma of the large arcade is heated to coronal temperatures and the loops become bright in the AIA~193 channel. As shown in \citet{2018A&A...619A..55M} and \citet{2019A&A...623A..78G}, CBPs can produce  a series of homologous eruptions (this CBP has produced at least six eruptions) while the CBP retains its general magnetic field configuration.  

The NLFFF modelling reproduces not only the overlying CBP loop system (see the left frame in Fig.~\ref{fig7}) but also 
a magnetic flux rope (right frame) located inside the CBP loop arcade. Both magnetic systems are shown overplotted on the H$\alpha$ image taken at the time of the eruption (16:13~UT). The location of the magnetic flux rope corresponds to the location of the MF in the H$\alpha$ images.
The left frame shows that the micro-flare kernels are associated with specific parts of the footpoints of the loop structure, while the flux rope lies between the micro-flare kernels. The flux rope has three clearly distinguishable footpoints, but only one footpoint appears to be rooted in the top left kernel (kernel A, see Fig.~\ref{fig2}) where some of the overlying coronal loops are also embedded. 
The structure of the overlying loop system is relatively simple, while the structure of the flux rope is rather complicated. 
The flux rope splits into two branches half way along its length, where there are both east-west and north-south orientated parts. The magnetic field lines in the north-south part of the  magnetic flux rope  join the east-west part and reverse in direction to connect towards the negative flux concentration that lies at the upper left of the image.  Exactly how this complicated field line structure of the flux rope is associated with the observed eruption around this time is unclear. This cannot be explained by our modelling approach that considers sequences of NLFFFs. 
It is interesting to note that there is a significant current along the north-south fraction of the flux rope, which may be the location of the flux rope destabilisation. However, from the NLFFF simulation it is clear that a highly non-potential magnetic field exists between the micro-flare kernels at the time of the eruption. This non-potential structure is created by the surface motions observed in the HMI magnetogram.

\begin{figure*}[!ht]
\centering
\hspace*{-0.95cm}
\includegraphics[scale=0.17]{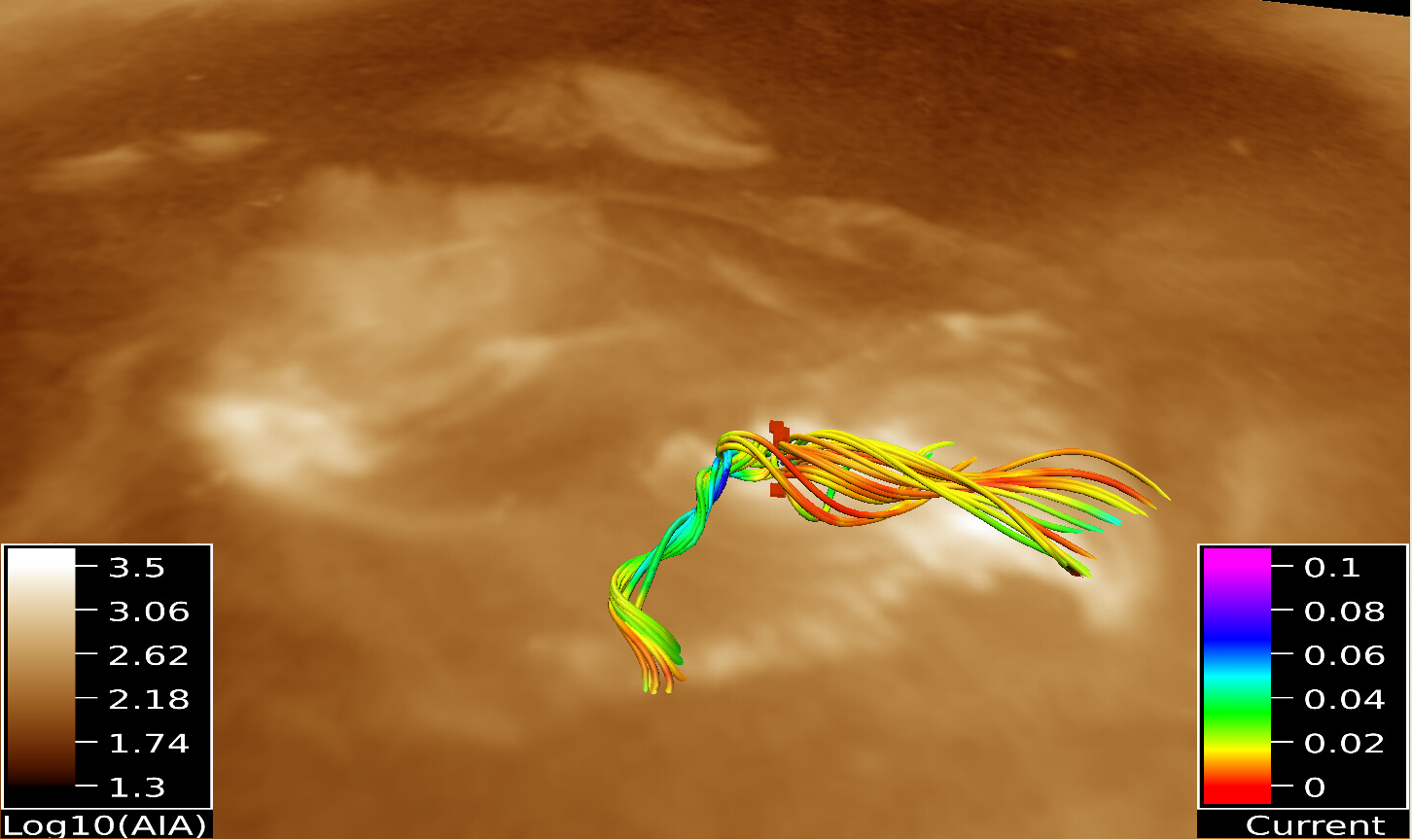} 
\includegraphics[scale=0.17]{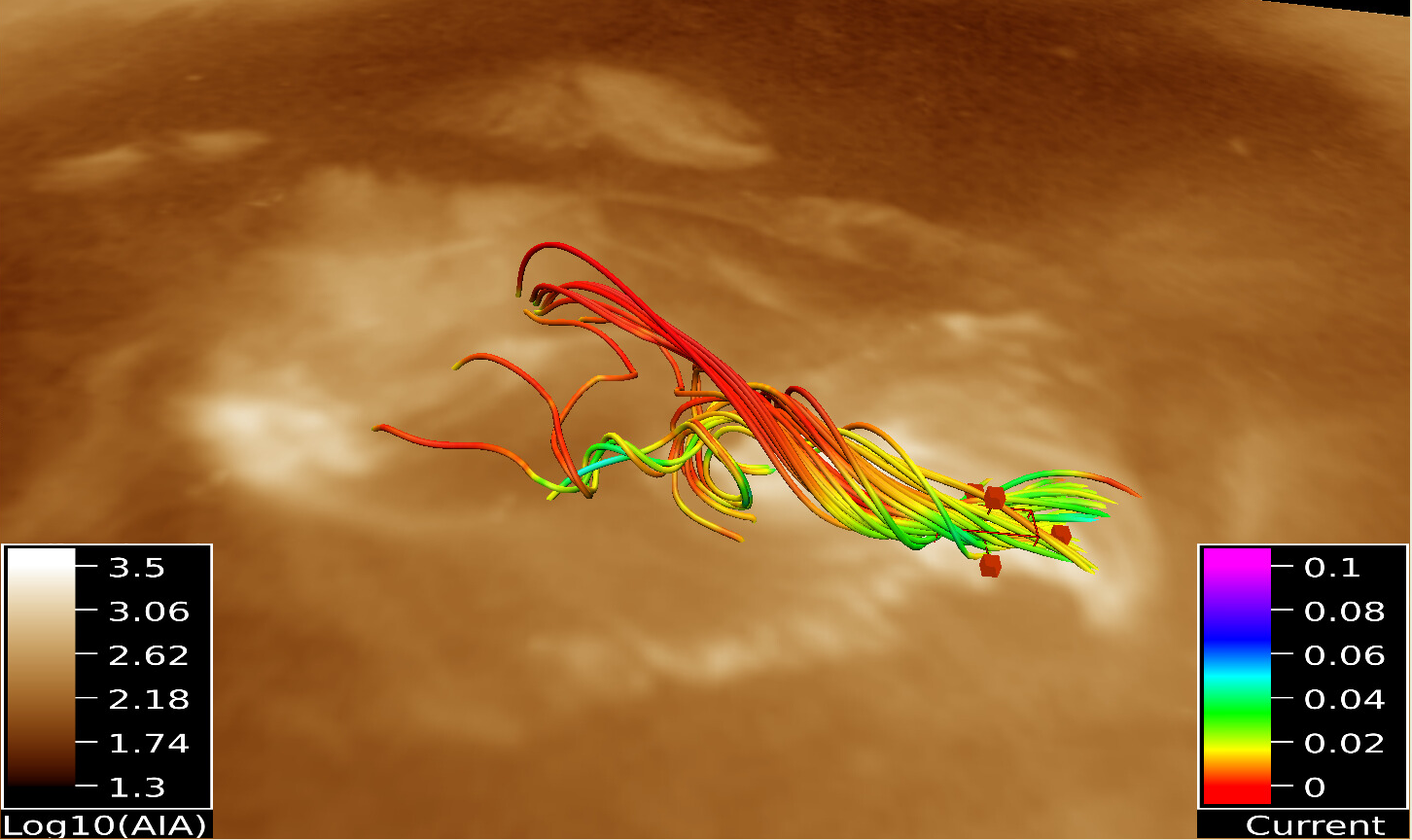}\\ 
\hspace*{-0.95cm}
\includegraphics[scale=0.17]{f8c_38287.png} 
\includegraphics[scale=0.17]{f8d_38287.png} 
\caption{ Magnetic field lines traced from two different source region for an HMI magnetogram taken at 16:15~UT. Left column: Magnetic field lines of the reconstructed brightest loop structure as traced from  the location of one of the micro-flare kernel (B). Right column: Magnetic field lines traced from the micro-flare kernel C kernel. The latter location corresponds to one of the footpoints of the filament flux rope. The red squares with connecting red lines show the starting location from which the magnetic field lines are traced. Top row background: log10(AIA~193~\AA) images taken at 16:15~UT. Bottom row background: HMI magnetograms taken at 16:15~UT.}
\label{fig8}
\end{figure*}

Figure~\ref{fig8} shows the local magnetic field line connectivity from the two flare kernel regions (B and C in Fig.\ref{fig2})  at the time during the  observed eruption. From kernel B  the connectivity is mainly towards kernel C, where there are mostly low lying field lines with only weak current. From kernel C the connectivity is more complex. Here it connects with both kernels B and A, where some of the field lines exhibit a higher twist even though they still only have a small amount of current along them. This clearly shows how complicated the general magnetic field structure.

\begin{figure*}[!ht]
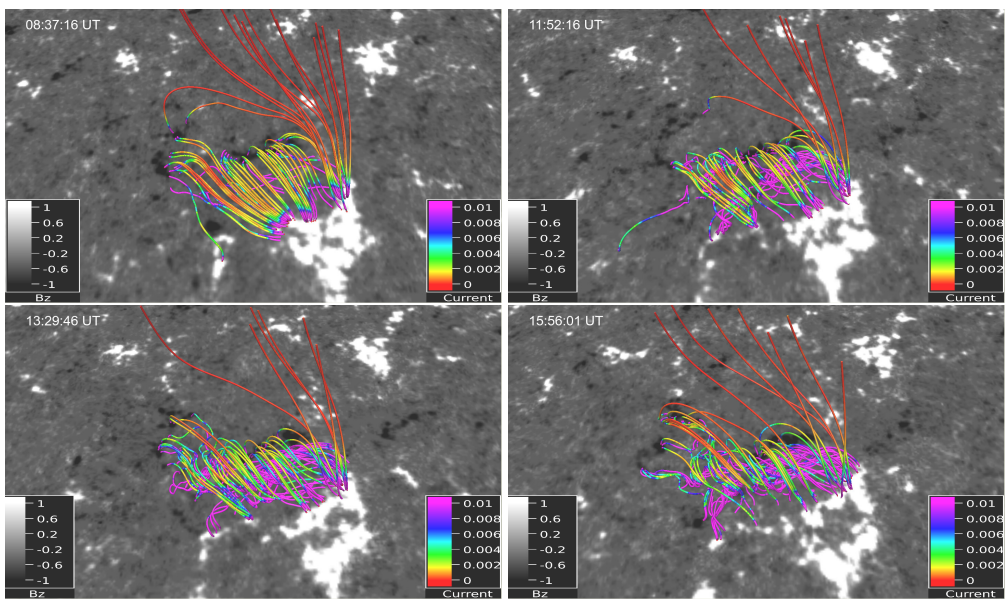

\centering
\hspace*{-0.95cm}
\includegraphics[scale=0.17]{f9a_38287.png}
\includegraphics[scale=0.17]{f9b_38287.png}\\
\hspace*{-0.95cm}
\includegraphics[scale=0.17]{f9c_38287.png}
\includegraphics[scale=0.17]{f9d_38287.png}
\caption{ Magnetic field lines traced from a large source region covering the region of  the positive and negative  magnetic flux  concentrations associated with the CBP. The frames show how a twisted flux structure builds up over time  in response to the imposed changes in the footpoint positions. An animation is also provided online, movie\_rope\_evol.mov.}
\label{fig9}
\end{figure*}

The NLFFF modelling approach used to simulate the 3D magnetic field structure allows us to follow the time dependent buildup of the twisted magnetic flux rope that forms above the PIL between the two polarity regions at the location of the observed MF.
In  Fig.~\ref{fig9}  we show four frames taken at 08:37~UT, 11:52~UT, 13:29~UT, and 15:56~UT that  show the buildup of the flux rope. We also provide an animation  of the image sequence for the time period between 08:37~UT and 16:35~UT (link in the caption of Fig.~\ref{fig9}). It is important to note that the initial condition of the simulation is a potential magnetic field.
Therefore an extended time period is required for the buildup of electric currents and sufficient free magnetic energy to  form
twisted structures above the PIL region. The frames show examples of this process, where as time progresses it can be seen that the twisted structure forms and increases in size where it is present well in advance of the actual eruption of the CBP region. Within the modelling there is no  eruption of the flux rope, instead the twisted region continues to increase in complexity as the modelling progresses in time. One possible reason for this is that this type of structure can be  seen to become numerically unstable when investigated for regions that are larger and therefore better resolved \citep{2014ApJ...782...71G,2013A&A...554A..77P,2014A&A...568A.120P}. The lack of instability here may therefore be partly due  to the small scale of the region which results in a relatively low resolution  of the region of interest, at the CBP. This does not allow the free magnetic energy to increase enough and the structure to become unstable leading to its disruption. However, it is clear that the surface motions deduced from the normal component magnetograms produce a highly nonpotential flux rope at the correct location of the MF.

\section{Discussion}
\label{disc}

Co-temporal observations of the chromosphere, transition region, and the hot (X-ray) solar corona 
combined with photospheric magnetic fields are essential for understanding the causes for solar eruptions and the processes that take place during any eruptive phenomenon. These observations combined with data-driven models can reveal, within certain limitations, the complexity of the physical processes that take place during eruptive phenomena. Such phenomenon but on a small-scale, originating from a coronal bright point in an equatorial coronal hole is the subject of the detailed analysis of the present paper. CBPs are highly important and suitable phenomena to study eruptions as they are well defined and spatially isolated from other structures and dynamic phenomena. This is in contrast to
the similar but far larger and more complex active regions.  
Thus, for CBP the identification of a trigger/cause as well as understanding of the spatial and time evolution of all involved plasma and magnetic structures is far easier  despite occasional spatial resolution issues. To eliminate these resolution limitations, the present study concentrated on the largest identified coronal hole CBP that produces a mass ejection consisting of a mini-filament eruption and an EUV/X-ray jet. While a series of studies have reported and discussed the role of mini-filaments as triggers of jets (in coronal holes) and mini-CMEs \citep[in the quiet Sun, e.g. see][and the references therein]{2018A&A...619A..55M}, data-driven modelling of their formation and investigations of the possible cause of their eruption has only been addressed by \cite{2019A&A...623A..78G}. On the other hand, quiet Sun and active region filament formation and eruption have been intensively studied both observationally and theoretically over several decades \citep[e.g.,][and the references therein]{2010SSRv..151..333M,2014LRSP...11....1P,2018LRSP...15....7G}.

Point-like brightenings have been reported in numerous studies of jets or CBP eruptions, and have been named micro-flares or jet bright points \citep[for details see the dedicated discussion in Section~4.5 of][]{2018A&A...619A..55M}. 
These events appear at the location of the filament eruption, i.e. at the polarity inversion line. Given the small scale of the mini-filaments and their associated polarity inversion line, the micro-flare is usually localised over just a few arcseconds. Therefore, its fine details  are hard to resolve including its relationship to the photospheric or coronal magnetic field configuration. Thanks to the large scale CBP analysed in the present study, we are able to investigate in full detail the nature of this point/like brightening or micro-flare. The micro-flare analysed here was clearly identified with three distinctive flare kernels that are known to be the signature of an intensely heated chromosphere following an energy deposition from the corona and the release of fast non-thermal particles. Although some studies have reported delays in the appearance of these micro-flares/jet-bright-points in EUV images, as explained in \citet{2018A&A...619A..55M}, this is caused by the rising cool mini-filament obscuring them due to the extinction of EUV emission in the cool plasma of the filament body. When viewed on a very small scale as in the cases of  jets and mini-CMEs, these kernels will only appear as an intense point-like brightening  in images that register emission  with  temperatures between 10\,000~K and 10~MK \citep[e.g.][]{2013ApJ...766..127Y}.

The timing of the micro-flare associated with three distinctive kernels seen in H$\alpha$ and EUV, 
combined with the timing of the brightening in the corona above the rising mini-filament, indicate that the micro-flare occurs due to magnetic reconnection between the rising flux rope of the 
mini-filament and the magnetic field of the overlying corona. This is supported by the NLFFF modelling which shows that at the location of the mini-filament a  flux rope structure exists and above it lie coronal loops.
Based on this interpretation, it is important to discuss the interpretation of jet bright points from the simulation of jets caused by mini-filament eruptions by \citet{2017Natur.544..452W}. The authors  show an example of the observation of a jet bright point. Postflare loops in their simulations (figure~2b in their paper) are associated with the jet bright point seen at the footpoints of an EUV jet shown in their extended data (figure 2). Postflare loops form during the gradual phase of solar flares or microflares (in the present case they formed around 10 min after the flare and lasted for several hours during the micro-flare gradual phase). In contrast, the jet bright points or micro-flares referred to here, appear during the lift off, of the filament.  They arise the moment the flux rope reaches the overlying loops and the energy release and deposition, e.g. magnetic reconnection, takes place. Thus jet bright points or micro-flares actually represent spatially unresolved  micro-flare kernels.

An NLFFF modelling of eruptive phenomena, such as the one presented here, does not allow  us to clearly follow the dynamical evolution across the eruption phase. Instead it only allows us to consider the slow quasi-static evolution of the magnetic field during the build-up phase of the eruption. From modelling of active regions,
\citet{2014ApJ...782...71G} and \citet{2018ApJ...852...82Y} found that the NLFFF modelling is able to evolve across eruptions, although it is not able to handle the dynamical evolution of the actual explosive event. To follow dynamic eruptive  events, one at least requires a full MHD simulation with both a realistic magnetic configuration and atmospheric model. The present modelling is able to clearly show that the initial potential  magnetic field can be evolved 
by surface motions alone into a NLFFF configuration that contains free magnetic energy in the region that observationally hosts the eruption. The comparison of the magnetic field structure 
to the observations shows that the NLFFF magnetic model contains many of the features that can explain the different observational signatures of the evolution and the eruption of the CBP. The model shows the presence of a complicated flux rope at the location where the observed mini-filament eruption is found to take place. Combining this with the information found in the previous investigation by \citet{2019A&A...623A..78G}, it is clear that the eruptions seen in the majority of CBPs require the presence of a flux rope in order for the eruption to take place. One open question is how realistic is the flux rope found in this time dependent NLFFF modelling compared to the real magnetic field configuration in the solar environment. The modelling here is based on the evolution of the normal magnetic field component. This is naturally a limitation compared to having access to the full magnetic field vector at the solar surface.  HMI vector magnetograms can only be used in regions with strong magnetic fields because of the low signal-to-noise of the transverse component of the field in the quiet Sun \citep{2014SoPh..289.3483H, 2009SoPh..260...83L}. The noise level of the HMI transverse component is 100~G, while the CBP transverse magnetic-field component are far below this value. Thus HMI vector magnetograms cannot be used for studying CBPs.
 To 
obtain such vector data requires the observational techniques to be improved on three 
fronts. First, the threshold for getting reliable vector magnetograms, second, a higher spatial  
resolution to better follow the small-scale structural changes, and third, a time resolution that makes it possible to follow the evolution in much better detail. Presently, only the Spectro-Polarimeter of the Solar Optical Telescope on board Hinode provides  vector field measurements at a sufficiently high signal-to-noise ratio. Unfortunately, these observations have  a limited field-of-view at a relatively low cadence (not better than 14~min) which is insufficient to constrain the present model.

\section{Summary and Conclusions}
\label{concl}

Given the estimated occurrence  rate of mini-CMEs and jets  to be at least 870 per day \citep{2018A&A...619A..55M} over the whole Sun, and their possible impact on the upper solar corona and solar wind, it is clear that understanding their trigger and evolution is of key importance. In the time of the two state-of-the-art space missions Solar Orbiter and Parker probe, combined with the largest ground based observatories such as the Daniel K. Inouye Solar Telescope and the European Solar Telescope, along with our ever improving theoretical modelling, we have a unique opportunity to advance our knowledge on these phenomena.

Here we present a case study of an eruption from a CBP located in an equatorial hole which is a continuation of our investigation of eruptions from CBPs (the previous were dedicated to CBPs in the QS). The following series of events occurred during the  eruptive phenomenon which can be deduced through studying both the observations and the NLFFF model. A mini-filament formed beneath the arcade of a large CBP located in a coronal hole around 3--4~h before the eruption, on 2013 October 12. The NLFFF modelling of the coronal field shows a flux rope forming at the location of the observed MF which is well aligned in time with the observations. During the formation of the MF (observations) or flux rope (NLFFF model) no significant photospheric magnetic flux concentration displacement (convergence) was observed and no significant magnetic flux cancellation between the two main magnetic polarities forming the CBP, was detected in the time leading to the dynamic phenomenon. The total unsigned flux did steadily decreases over 10~hr (see Fig.~\ref{app1}) but this is the typical evolution of magnetic flux associated with any CBP whether or not they produce coronal jets or mini-CMEs. Therefore, the flux rope formation in the model at the spatial location of the MF must have been generated by small-scale footpoint motions and cancellation.  A  micro-flare that occurred at the liftoff of the MF  is associated with three flare kernels that formed shortly after the MF liftoff. No observational signature is found for reconnection beneath the erupting MF. 

The applied NLFFF modelling has clearly demonstrated that an initial potential field can be evolved into a non-potential magnetic field configuration that contains free magnetic energy in the region that observationally hosts the eruption.  The comparison of the magnetic field structure revealed that the magnetic NLFFF model contains many of the features that can explain the different observational signatures found in the evolution and eruption of the mini-filament and the CBP. In future such modelling may eventually indicate the location of destabilisation that results in
the eruptions of flux ropes.

\begin{appendix}
\section{Data driven NLFFF simulations}
\label{sect:method}

To simulate the 3D coronal evolution of the CBP from the magnetogram observations a time-dependent NLFFF relaxation technique is applied \citep{2014ApJ...782...71G}. This technique follows the evolution of the 3D magnetic field of the CBP and surrounding regions, where the evolution of the coronal field is directly driven by the magnetogram data. 
The NLFFF relaxation technique may be used to simulate the solar corona, as in the corona the  Alfv\'en speed is approximately one order of magnitude greater than the sound speed. This means that magnetic forces are dominant over plasma forces and magnetic fields are in a force-free state. While there are three  force-free assumptions that can be made (potential, linear and non-linear) we choose to model the magnetic field of the CBP using the non-linear force-free assumption as it is the most realistic. With this modelling technique we may simulate the slow quasi-static evolution of the solar corona through a series of non-linear force-free states where these states are a consequence of the boundary evolution obtained from the observed magnetograms. It is important to note that the non-linear force-fee modelling technique and approximation applied, is only valid for near equilibrium coronal conditions in the absence of eruptions. Due to this it can be used to model the build up of stress and free magnetic energy to the point of an eruption, but not the eruption itself. Once an eruption occurs rapid dynamics take place and pressure forces can no longer be neglected 
\citep{2013A&A...554A..77P,2014A&A...568A.120P}. The technique can however be used to understand the pre-eruptive and build-up to eruption magnetic field configurations. To apply this technique a number of stages need to be applied from data preparation, construction of the initial condition and finally the full simulation where each of these stages are now described in detail. 

The data preparation stage produces a long time series of normal component magnetograms that are used as the evolving
boundary condition at the photosphere which drives the 3D coronal magnetic field and simulation. Full details of this process can be found in 
Paper~1 where we briefly recap the process. First the time resolution of the HMI time series is reduced from 45\,s to
450\,s. This is carried out to eliminate any high-frequency noise that exists between two consecutive magnetograms
and allows for a more clean determination of the systematic time changes of the magnetic features between subsequent magnetograms. In addition to this temporal change to the time series of magnetograms an additional spatial clean up and smoothing is applied to the 2D HMI magnetograms.  
This included the removal of single-pixel clusters with unrealistically high count values. In contrast to previous studies which considered active regions \citep{2014ApJ...782...71G, 2018ApJ...852...82Y} no lower flux threshold for zeroing pixels values was adopted. This is because the flux regions under investigation are small in pixel sizes compared to active regions and removing flux below a given threshold may strongly influence the magnetic field topology and the derived time evolution. Once these processes are carried out a long time series of magnetogram data representing the evolution of the magnetic field in the photosphere underneath and surrounding the CBP is produced. These data show typical features occurring in the magnetic carpet including emergence, coalescence, fragmentation, and cancellation. The time series is made sufficiently
long such that the slow systematic changes in the photospheric magnetic field configuration that systematically stress the 3D coronal magnetic field can be followed over time. 

The continuous time evolution of the magnetic field obtained through the applied magnetograms is assumed to be 2D periodic in the horizontal direction. For the NLFFF simulations the primary variable is the magnetic vector potential, $\bf{A}$. To simulate the CBP a time series of vector potentials are derived at the photosphere based on the normal component magnetograms. To change the magnetic field on the photospheric boundary in accordance with the observations, it is assumed that the two horizontal components of the vector potential in the photospheric plane can be represented by a scalar potential ($\Phi$) in the following way,
\EQA
A_x &=&  {\frac{\partial \Phi}{\partial y}},\\
A_y &=& -{\frac{\partial \Phi}{\partial x}}.
\ENA
Using the general definition of the magnetic field by a vector potential, $\bf{B} = \nabla \times \bf{A}$, and setting the gauge to zero, these two approaches are combined to provide a Poisson equation for determining the scalar potential $\Phi$ based on the knowledge of the magnetic field at the bottom boundary,
\EQ
{\frac{\partial^2 \Phi}{\partial x^2}} + {\frac{\partial^2 \Phi}{\partial y^2}} = - B_z.
\EN
Assuming the data in the 2D plane are periodic, this equation is solved using a FFT approach where $B_z$ represent the normal field component. 
Once the time series of the magnetogram data is produced the first frame is used to construct an initial potential field.
As none of the magnetograms are in a perfect flux balance, the top boundary of the domain is open which allows the excess magnetic flux to exit. Again to construct the initial potential 3D magnetic field a fast Fourier transform (FFT) approach is used where this solution may be expanded in height defining an initial potential magnetic field using the Devore Gauge \citep{2000ApJ...539..944D}.

To simulate the coupled evolution of the photospheric and coronal magnetic fields through a continuous sequence of NLFFF  solutions driven by the evolution of the corrected HMI magnetograms at the photosphere, the following technique is applied. To start the simulation the vector potential {\bf$A$} describing the initial potential field is taken along with its deduced coronal field. Subsequently, the vector potential components at the base $(A_x, A_y)$ are updated, resulting in the time evolution of the normal magnetic field at the photosphere from the present observed magnetogram to the next. The effect of this boundary evolution is to inject electric currents, a Poynting flux and non-potentiality into the coronal field which evolves the coronal field away from equilibrium. In response to this, the vector potential in the full 3D domain is found by solving the uncurled induction equation, 
\EQ
{\frac{\partial {\bf A}}{\partial t}} = {\bf v} \times {\bf B} + {\bf R_{num}}
,\EN
where ${\bf v}$ is the magneto-frictional velocity, expressed by
\EQ
{\bf v} = {\frac{1}{\nu}} {\frac{{\bf j} \times {\bf B}}{B^2}},
\EN
and ${\bf R_{num}}$ is a non-ideal term that allows for numerical diffusion. The role of the magneto-frictional velocity is to return the coronal field to an equilibrium force-free state -- in general a non-linear force-free field. 
For each update of the boundary conditions, provided by the corrected HMI data, the induction equation is solved in a frictional time until the magneto-frictional velocity becomes sufficiently low. This indicates that a new near NLFFF state has been reached and a snapshot of the 3D vector potential, $\bf A$, is saved. Using this technique a continuous time sequence of NLFFF can be produced from the observed magnetograms. A full description of the code is given in \citet{2011ApJ...729...97M} and \citet{2014ApJ...782...71G}.

It is important to note that for the non-linear force-free modelling technique that we apply in this paper we do not use or require vector field information at the photosphere. As mentioned in Section 3.2  HMI vector magnetograms are only suitable for regions with strong magnetic fields because of the low signal-to-noise ratio of the transverse component of the field in the quiet Sun \citep{2014SoPh..289.3483H,2009SoPh..260...83L}. The noise level of the HMI transverse component is 100 G \citep[e.g.][]{2013A&A...550A..14T}, while the CBP transverse component is far below this value. Thus HMI vector magnetograms cannot be used for studying CBPs.  Therefore the technique we apply is very useful when vector data are not sufficient to constrain the horizontal field at the photosphere. Rather any non-potential horizontal fields are self-consistently produced due to the applied evolution of the normal field component and its subsequent Poynting flux injection into the corona.
Once constructed the  3D vector field from the simulation is analysed using  VAPOR in an attempt to better understand the structural evolution of the magnetic field, with an emphasis on the region around the erupting CBP.

\section{Online material}
\begin{figure}[!ht]
\centering
\hspace*{-1cm}
\includegraphics[scale=0.50]{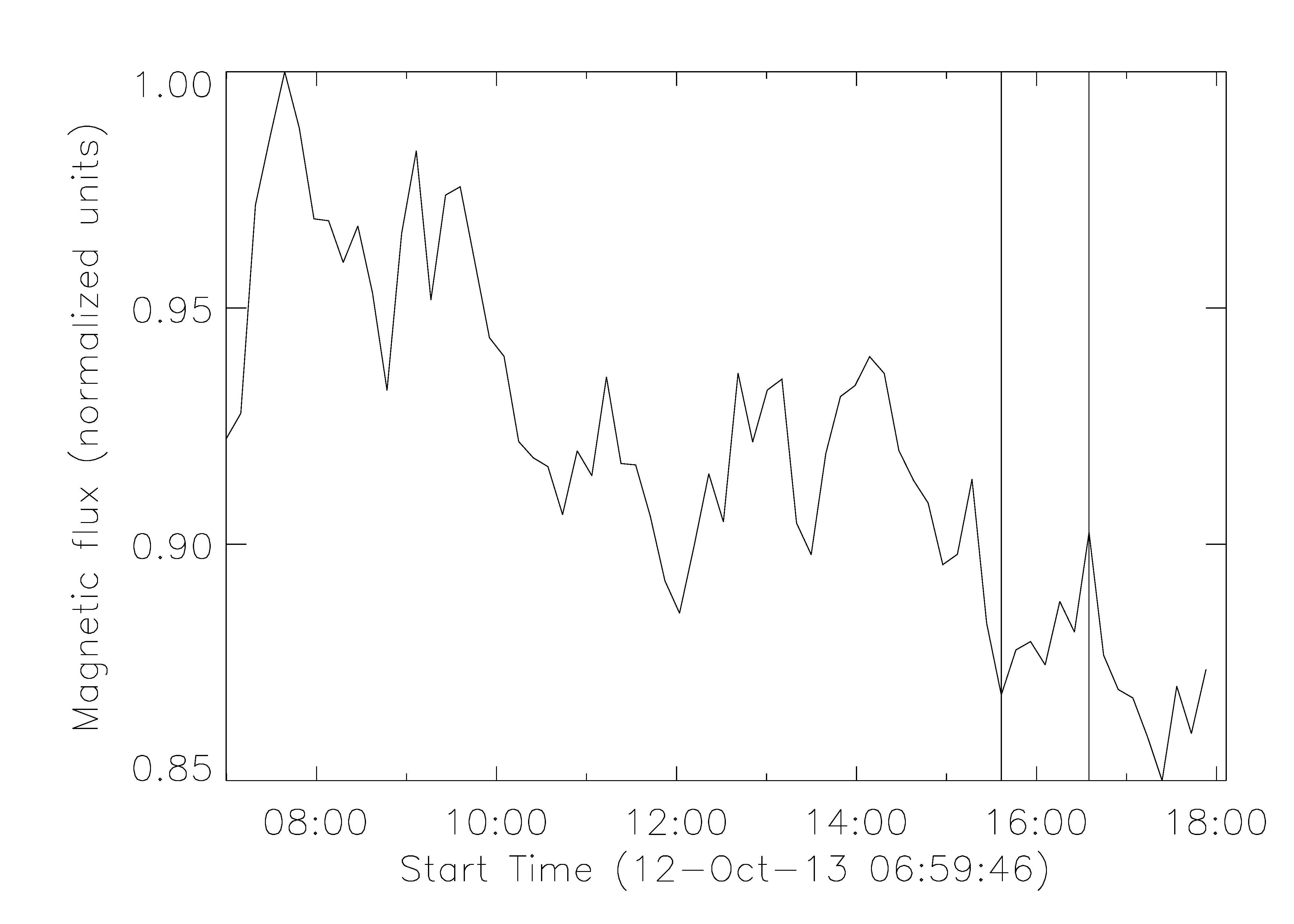}
\caption{Normalized total unsigned magnetic flux temporal evolution from the region outlined with a white square on the HMI panel in Fig.~\ref{fig2} at 16:35:01~UT covering the time interval from 07:00 to 17:53~UT on 2013 October 12. The vertical lines indicate the period of the eruption.}
\label{app1}
\end{figure}

\begin{figure}[!ht]
\centering
\hspace*{-1cm}
\includegraphics[scale=0.15]{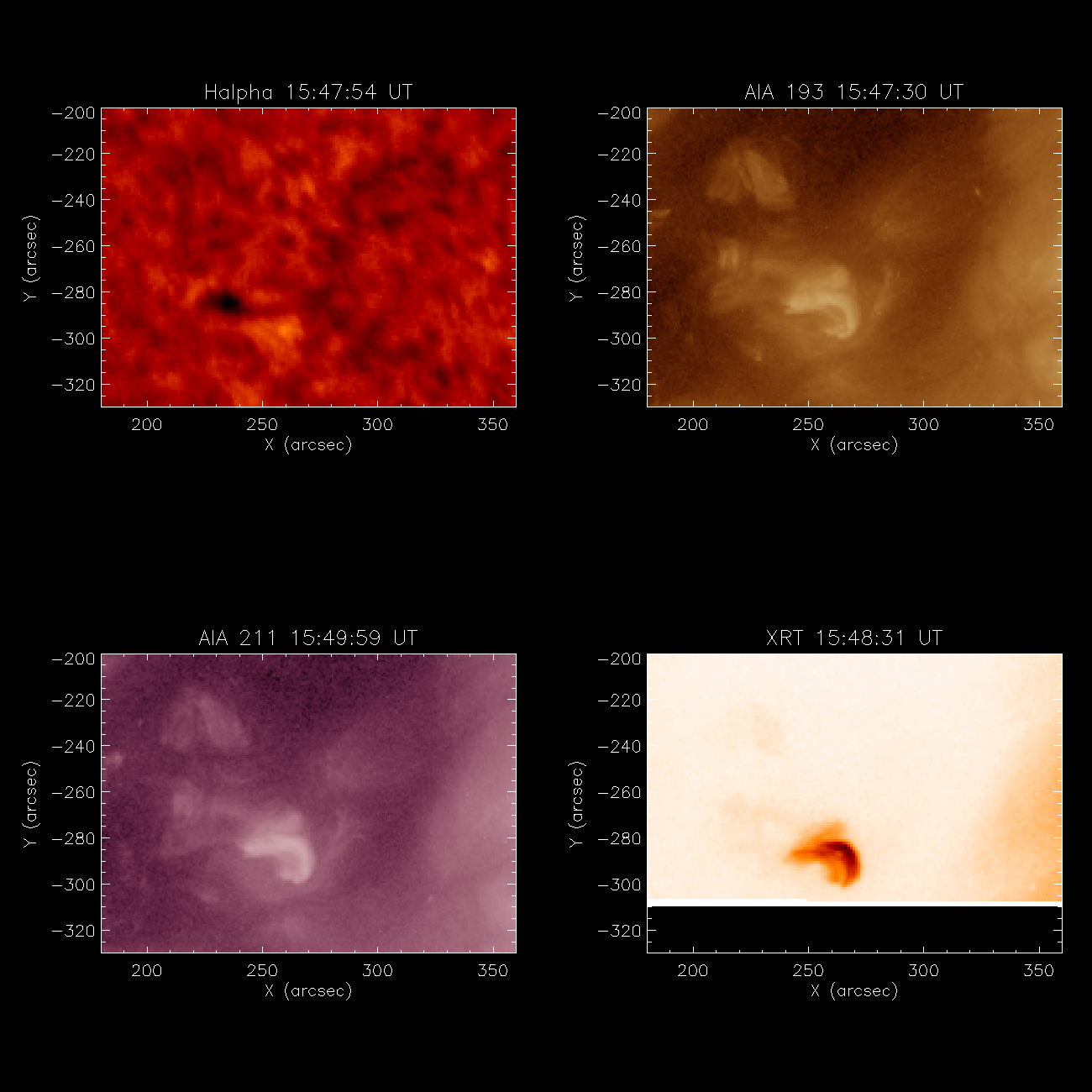}
\caption{Animation of a sequence of GONG H$\alpha$, AIA~193 (top row), AIA~211, and XRT  (bottom row)  images showing the series of events including mini-filament eruption and EUV/X-ray jet formation, as well as the flare ribbon and postflare loop formation.}
\label{app2}
\end{figure}

\end{appendix}

\begin{acknowledgements}

We would like to thank the anonymous referee for their helpful comments and suggestions. MM was supported by the Brain Pool Program of the National Research Foundation of Korea (NRF-2019H1D3A2A01099143). DHM would like to thank the Science and Technology Facilities Council (UK) through the consolidated grant ST/N000609/1 and the European Research Council (ERC) under the European Union Horizon 2020 research and innovation program (grant agreement No. 647214). VAPOR is a product of the National Center for Atmospheric Research's Computational and Information Systems Lab. Support for VAPOR is provided by the U.S. National Science Foundation (grants $\#$ 03-25934 and 09-06379, ACI-14-40412), and by the Korea Institute of Science and Technology Information.
The HMI data are provided courtesy of NASA/SDO and corresponding science teams. The HMI data have been retrieved using the Stanford University's Joint Science Operations Centre/Science Data Processing Facility. M.M. and K.G. thank the ISSI Bern for the support to the team ``Observation-Driven Modelling of Solar Phenomena''. 

\end{acknowledgements}

\bibliographystyle{aa}

\begin{thebibliography}{32}
\expandafter\ifx\csname natexlab\endcsname\relax\def\natexlab#1{#1}\fi

\bibitem[{{DeVore}(2000)}]{2000ApJ...539..944D}
{DeVore}, C.~R. 2000, \apj, 539, 944

\bibitem[{{Fisher} {et~al.}(1985){Fisher}, {Canfield}, \&
  {McClymont}}]{1985ApJ...289..414F}
{Fisher}, G.~H., {Canfield}, R.~C., \& {McClymont}, A.~N. 1985, \apj, 289, 414

\bibitem[{{Fletcher}(2012)}]{2012ASPC..456..183F}
{Fletcher}, L. 2012, Astronomical Society of the Pacific Conference Series,
  Vol. 456, {Solar Chromospheric Flares: Energy Release, Transport and
  Radiation}, ed. L.~{Golub}, I.~{De Moortel}, \& T.~{Shimizu}, 183

\bibitem[{{Fletcher} {et~al.}(2011){Fletcher}, {Dennis}, {Hudson}, {Krucker},
  {Phillips}, {Veronig}, {Battaglia}, {Bone}, {Caspi}, {Chen}, {Gallagher},
  {Grigis}, {Ji}, {Liu}, {Milligan}, \& {Temmer}}]{2011SSRv..159...19F}
{Fletcher}, L., {Dennis}, B.~R., {Hudson}, H.~S., {et~al.} 2011, \ssr, 159, 19

\bibitem[{{Galsgaard} {et~al.}(2019){Galsgaard}, {Madjarska}, {Mackay}, \&
  {Mou}}]{2019A&A...623A..78G}
{Galsgaard}, K., {Madjarska}, M.~S., {Mackay}, D.~H., \& {Mou}, C. 2019, \aap,
  623, A78

\bibitem[{{Gibb} {et~al.}(2014){Gibb}, {Mackay}, {Green}, \&
  {Meyer}}]{2014ApJ...782...71G}
{Gibb}, G.~P.~S., {Mackay}, D.~H., {Green}, L.~M., \& {Meyer}, K.~A. 2014,
  \apj, 782, 71

\bibitem[{{Gibson}(2018)}]{2018LRSP...15....7G}
{Gibson}, S.~E. 2018, Living Reviews in Solar Physics, 15, 7

\bibitem[{{Golub} {et~al.}(2007){Golub}, {Deluca}, {Austin}, {Bookbinder},
  {Caldwell}, {Cheimets}, {Cirtain}, {Cosmo}, {Reid}, {Sette}, {Weber},
  {Sakao}, {Kano}, {Shibasaki}, {Hara}, {Tsuneta}, {Kumagai}, {Tamura},
  {Shimojo}, {McCracken}, {Carpenter}, {Haight}, {Siler}, {Wright}, {Tucker},
  {Rutledge}, {Barbera}, {Peres}, \& {Varisco}}]{2007SoPh..243...63G}
{Golub}, L., {Deluca}, E., {Austin}, G., {et~al.} 2007, \solphys, 243, 63

\bibitem[{{Golub} {et~al.}(1974){Golub}, {Krieger}, {Silk}, {Timothy}, \&
  {Vaiana}}]{1974ApJ...189L..93G}
{Golub}, L., {Krieger}, A.~S., {Silk}, J.~K., {Timothy}, A.~F., \& {Vaiana},
  G.~S. 1974, \apjl, 189, L93

\bibitem[{{Hermans} \& {Martin}(1986)}]{1986NASCP2442..369H}
{Hermans}, L.~M. \& {Martin}, S.~F. 1986, in NASA Conference Publication, Vol.
  2442, Coronal and Prominence Plasmas, ed. A.~I. {Poland} (NASA), 369--375

\bibitem[{{Hoeksema} {et~al.}(2014){Hoeksema}, {Liu}, {Hayashi}, {Sun},
  {Schou}, {Couvidat}, {Norton}, {Bobra}, {Centeno}, {Leka}, {Barnes}, \&
  {Turmon}}]{2014SoPh..289.3483H}
{Hoeksema}, J.~T., {Liu}, Y., {Hayashi}, K., {et~al.} 2014, \solphys, 289, 3483

\bibitem[{{Hudson}(2007)}]{2007ASPC..368..365H}
{Hudson}, H.~S. 2007, Astronomical Society of the Pacific Conference Series,
  Vol. 368, {Chromospheric Flares}, ed. P.~{Heinzel}, I.~{Dorotovi{\v{c}}}, \&
  R.~J. {Rutten}, 365

\bibitem[{{Innes} {et~al.}(2009){Innes}, {Genetelli}, {Attie}, \&
  {Potts}}]{2009A&A...495..319I}
{Innes}, D.~E., {Genetelli}, A., {Attie}, R., \& {Potts}, H.~E. 2009, \aap,
  495, 319

\bibitem[{{Innes} {et~al.}(2010){Innes}, {McIntosh}, \&
  {Pietarila}}]{2010A&A...517L...7I}
{Innes}, D.~E., {McIntosh}, S.~W., \& {Pietarila}, A. 2010, \aap, 517, L7

\bibitem[{{Kamio} {et~al.}(2011){Kamio}, {Curdt}, {Teriaca}, \&
  {Innes}}]{2011A&A...529A..21K}
{Kamio}, S., {Curdt}, W., {Teriaca}, L., \& {Innes}, D.~E. 2011, \aap, 529, A21

\bibitem[{{Leka} {et~al.}(2009){Leka}, {Barnes}, {Crouch}, {Metcalf}, {Gary},
  {Jing}, \& {Liu}}]{2009SoPh..260...83L}
{Leka}, K.~D., {Barnes}, G., {Crouch}, A.~D., {et~al.} 2009, \solphys, 260, 83

\bibitem[{{Lemen} {et~al.}(2012){Lemen}, {Title}, {Akin}, {Boerner}, {Chou},
  {Drake}, {Duncan}, {Edwards}, {Friedlaender}, {Heyman}, {Hurlburt}, {Katz},
  {Kushner}, {Levay}, {Lindgren}, {Mathur}, {McFeaters}, {Mitchell}, {Rehse},
  {Schrijver}, {Springer}, {Stern}, {Tarbell}, {Wuelser}, {Wolfson}, {Yanari},
  {Bookbinder}, {Cheimets}, {Caldwell}, {Deluca}, {Gates}, {Golub}, {Park},
  {Podgorski}, {Bush}, {Scherrer}, {Gummin}, {Smith}, {Auker}, {Jerram},
  {Pool}, {Soufli}, {Windt}, {Beardsley}, {Clapp}, {Lang}, \&
  {Waltham}}]{2012SoPh..275...17L}
{Lemen}, J.~R., {Title}, A.~M., {Akin}, D.~J., {et~al.} 2012, \solphys, 275, 17

\bibitem[{{Mackay} {et~al.}(2011){Mackay}, {Green}, \& {van
  Ballegooijen}}]{2011ApJ...729...97M}
{Mackay}, D.~H., {Green}, L.~M., \& {van Ballegooijen}, A. 2011, \apj, 729, 97

\bibitem[{{Mackay} {et~al.}(2010){Mackay}, {Karpen}, {Ballester}, {Schmieder},
  \& {Aulanier}}]{2010SSRv..151..333M}
{Mackay}, D.~H., {Karpen}, J.~T., {Ballester}, J.~L., {Schmieder}, B., \&
  {Aulanier}, G. 2010, \ssr, 151, 333

\bibitem[{{Madjarska}(2019)}]{2019LRSP...16....2M}
{Madjarska}, M.~S. 2019, Living Reviews in Solar Physics, 16, 2

\bibitem[{{Mou} {et~al.}(2018){Mou}, {Madjarska}, {Galsgaard}, \&
  {Xia}}]{2018A&A...619A..55M}
{Mou}, C., {Madjarska}, M.~S., {Galsgaard}, K., \& {Xia}, L. 2018, \aap, 619,
  A55

\bibitem[{{Pagano} {et~al.}(2013){Pagano}, {Mackay}, \&
  {Poedts}}]{2013A&A...554A..77P}
{Pagano}, P., {Mackay}, D.~H., \& {Poedts}, S. 2013, \aap, 554, A77

\bibitem[{{Pagano} {et~al.}(2014){Pagano}, {Mackay}, \&
  {Poedts}}]{2014A&A...568A.120P}
{Pagano}, P., {Mackay}, D.~H., \& {Poedts}, S. 2014, \aap, 568, A120

\bibitem[{{Parenti}(2014)}]{2014LRSP...11....1P}
{Parenti}, S. 2014, Living Reviews in Solar Physics, 11, 1

\bibitem[{{Pesnell} {et~al.}(2012){Pesnell}, {Thompson}, \&
  {Chamberlin}}]{2012SoPh..275....3P}
{Pesnell}, W.~D., {Thompson}, B.~J., \& {Chamberlin}, P.~C. 2012, \solphys,
  275, 3

\bibitem[{{Raouafi} {et~al.}(2016){Raouafi}, {Patsourakos}, {Pariat}, {Young},
  {Sterling}, {Savcheva}, {Shimojo}, {Moreno-Insertis}, {DeVore}, {Archontis},
  {T{\"o}r{\"o}k}, {Mason}, {Curdt}, {Meyer}, {Dalmasse}, \&
  {Matsui}}]{2016SSRv..201....1R}
{Raouafi}, N.~E., {Patsourakos}, S., {Pariat}, E., {et~al.} 2016, \ssr, 201, 1

\bibitem[{{Scherrer} {et~al.}(2012){Scherrer}, {Schou}, {Bush}, {Kosovichev},
  {Bogart}, {Hoeksema}, {Liu}, {Duvall}, {Zhao}, {Title}, {Schrijver},
  {Tarbell}, \& {Tomczyk}}]{2012SoPh..275..207S}
{Scherrer}, P.~H., {Schou}, J., {Bush}, R.~I., {et~al.} 2012, \solphys, 275,
  207

\bibitem[{{Tadesse} {et~al.}(2013){Tadesse}, {Wiegelmann}, {Inhester},
  {MacNeice}, {Pevtsov}, \& {Sun}}]{2013A&A...550A..14T}
{Tadesse}, T., {Wiegelmann}, T., {Inhester}, B., {et~al.} 2013, \aap, 550, A14

\bibitem[{{Wyper} {et~al.}(2017){Wyper}, {Antiochos}, \&
  {DeVore}}]{2017Natur.544..452W}
{Wyper}, P.~F., {Antiochos}, S.~K., \& {DeVore}, C.~R. 2017, \nat, 544, 452

\bibitem[{{Wyper} {et~al.}(2018){Wyper}, {DeVore}, {Karpen}, {Antiochos}, \&
  {Yeates}}]{2018ApJ...864..165W}
{Wyper}, P.~F., {DeVore}, C.~R., {Karpen}, J.~T., {Antiochos}, S.~K., \&
  {Yeates}, A.~R. 2018, \apj, 864, 165

\bibitem[{{Yardley} {et~al.}(2018){Yardley}, {Mackay}, \&
  {Green}}]{2018ApJ...852...82Y}
{Yardley}, S.~L., {Mackay}, D.~H., \& {Green}, L.~M. 2018, \apj, 852, 82

\bibitem[{{Young} {et~al.}(2013){Young}, {Doschek}, {Warren}, \&
  {Hara}}]{2013ApJ...766..127Y}
{Young}, P.~R., {Doschek}, G.~A., {Warren}, H.~P., \& {Hara}, H. 2013, \apj,
  766, 127

\end{thebibliography}

\end{document}